\documentclass[prd,nofootinbib,showpacs,preprint]{revtex4}
\usepackage{amsmath}
\usepackage{graphicx}

\usepackage{epsfig}
\graphicspath{{Figs/}}
\usepackage{dcolumn}
\usepackage{bm}
\usepackage{amssymb}
\usepackage[usenames,dvipsnames]{color}
\usepackage{slashed}
\usepackage[dvipdfm,colorlinks,citecolor=blue]{hyperref}
\begin{document}
\title{Constraining Type I Seesaw with $A_4$ Flavor Symmetry From Neutrino Data and Leptogenesis}
\author{Rupam Kalita}
\email{rup@tezu.ernet.in}
\author{Debasish Borah}
\email{dborah@tezu.ernet.in}
\affiliation{Department of Physics, Tezpur University, Tezpur-784028, India}


\begin{abstract}
We study a type I seesaw model of neutrino masses within the framework of $A_4$ flavor symmetry. Incorporating the presence of both singlet and triplet flavons under $A_4$ symmetry, we construct the leptonic mass matrices involved in type I seesaw mechanism. We then construct the light neutrino mass matrix using the $3\sigma$ values of neutrino oscillation parameters keeping the presently undetermined parameters namely, the lightest neutrino mass $m_{\text{lightest}}$, one Dirac CP phase $\delta$ and two Majorana phases $\alpha, \beta$ as free parameters. Comparing the mass matrices derived using $A_4$ parameters as well as light neutrino parameters, we then evaluate all the $A_4$ parameters in terms of light neutrino parameters.  Assuming some specific vacuum alignments of $A_4$ triplet flavon field, we then numerically evaluate all the free parameters in the light neutrino sector, using which we also find out the remaining $A_4$ parameters. We then use the numerical values of these parameters to calculate baryon asymmetry through the mechanism of leptogenesis. We not only constrain the $A_4$ vacuum alignments from the requirement of successful leptogenesis, but also constrain the free parameters in the light neutrino sector $(m_{\text{lightest}}, \delta, \alpha, \beta)$ to certain range of values. These values can be tested in ongoing and future neutrino experiments providing a way to discriminate between different possible $A_4$ vacuum alignments discussed in this work.

\end{abstract}
\maketitle

\section{Introduction}
Understanding the dynamical origin of fermion masses and mixing has been one of the most widely investigated problems in particle physics during last few decades. Although the standard model (SM) of particle physics can explain the origin of mass through the Higgs mechanism, it can not provide a justification to the observed fermion mass hierarchy and mixing. Any attempt to gain an insight into such problems inevitably requires input from beyond standard model (BSM) physics. The observed patterns of quark masses and mixing still remains a puzzle and a significant number of research works have been done in order to understand their fundamental origin. The leptonic mass and mixing, after the discovery of tiny neutrino masses and their large mixing \cite{PDG} have made the fermion mass and mixing problem even more puzzling. This is because the neutrino mass is found to lie at least twelve order of magnitude lower than the electroweak scale, and the pattern of leptonic mixing with large mixing angles is very different from quark mixing with small mixing angles. Recent neutrino experiments T2K \cite{T2K}, Double ChooZ \cite{chooz}, Daya-Bay \cite{daya} and RENO \cite{reno} have confirmed the earlier observations of tiny neutrino mass and large leptonic mixing and also measured the mixing parameters with more precision.  Apart from more precise measurements of neutrino parameters, these experiments also provided strong evidence for a non-zero value of reactor mixing angle $\theta_{13}$ which was thought to be (very close to) zero before. The best fit values of neutrino oscillation parameters that have appeared in the recent analysis of \cite{schwetz14}  and \cite{valle14}  are shown in table \ref{tab:data1} and \ref{tab:data2} respectively.
\begin{center}
\begin{table}[htb]
\begin{tabular}{|c|c|c|}
\hline
Parameters & Normal Hierarchy (NH) & Inverted Hierarchy (IH) \\
\hline
$ \frac{\Delta m_{21}^2}{10^{-5} \text{eV}^2}$ & $7.50$ & $7.50 $ \\
$ \frac{|\Delta m_{31}^2|}{10^{-3} \text{eV}^2}$ & $2.457$ & $2.449$ \\
$ \sin^2\theta_{12} $ &  $0.304$ & $0.304 $ \\
$ \sin^2\theta_{23} $ & $0.452$ &  $0.579 $ \\
$\sin^2\theta_{13} $ & $0.0218$ & $0.0219 $ \\
$ \delta_{CP} $ & $306^o$ & $254^o$ \\
\hline
\end{tabular}
\caption{Global best fit values of neutrino oscillation parameters \cite{schwetz14}}
\label{tab:data1}
\end{table}
\end{center}
\begin{center}
\begin{table}[htb]
\begin{tabular}{|c|c|c|}
\hline
Parameters & Normal Hierarchy (NH) & Inverted Hierarchy (IH) \\
\hline
$ \frac{\Delta m_{21}^2}{10^{-5} \text{eV}^2}$ & $7.60$ & $7.60$ \\
$ \frac{|\Delta m_{31}^2|}{10^{-3} \text{eV}^2}$ & $2.48$ & $2.38 $ \\
$ \sin^2\theta_{12} $ &  $0.323$ & $0.323 $ \\
$ \sin^2\theta_{23} $ & $0.567$ &  $0.573 $ \\
$\sin^2\theta_{13} $ & $0.0234$ & $0.0240 $ \\
$ \delta_{CP} $ & $254^o$ & $266^o$ \\
\hline
\end{tabular}
\caption{Global best fit values of neutrino oscillation parameters \cite{valle14}}
\label{tab:data2}
\end{table}
\end{center}

Due to the absence of right handed neutrino, the SM can not explain neutrino mass through the conventional Higgs mechanism. Even if the right handed neutrinos are added by hand to the SM in order to allow a Dirac mass term, then the respective Yukawa couplings have to fine tuned to the level of $10^{-12}$, which is highly unnatural. On the other hand, several BSM frameworks provide a natural origin of tiny neutrino mass by the conventional seesaw mechanism which can be broadly categorized into three types: type I \cite{ti}, type II \cite{tii} and type III \cite{tiii}, all of which involve the introduction of additional heavy fermion or scalar particles into the SM. Similarly, several BSM frameworks have also been proposed in order to generate large leptonic mixing. Most of these frameworks introduce additional flavor symmetries, either discrete or continuous, into the SM. Prior to the discovery of non-zero $\theta_{13}$, the neutrino oscillation data were in perfect agreement with some versions of $\mu-\tau$ symmetric neutrino mass matrix. Out of four different neutrino mixing patterns that can originate from such a $\mu-\tau$ symmetric neutrino mass matrix, the Tri-Bimaximal (TBM) \cite{Harrison} form of neutrino mixing received more attention in the literature. This particular mixing predicts the mixing angles as $\theta_{23} = 45^o, \theta_{12}=35.3^o, \theta_{13} = 0$. However, since $\theta_{13}=0$ has been ruled out by latest experimental data, one has to modify these $\mu-\tau$ symmetric or TBM type mass matrices in order to generate non-zero $\theta_{13}$. Since the measured value of $\theta_{13}$ is much smaller compared to the other two mixing angles, one can still consider TBM mixing to be valid at leading order and can explain non-zero $\theta_{13}$ by incorporating small perturbations. This has led to several works including \cite{nzt13, nzt13A4, nzt13GA,db-t2, dbijmpa, dbmkdsp, dbrk} within the framework of different BSM frameworks. 

The TBM type mixing can be accommodated within several discrete flavor symmetry models \cite{discreteRev}. Among them, the discrete group $A_4$, group of even permutations of four objects, can reproduce the TBM mixing in the most economical way \cite{A4TBM}. Necessary deviations from TBM mixing in order to generate non-zero $\theta_{13}$ can also be explained within the $A_4$ model as shown by many groups including \cite{nzt13A4, nzt13GA}. Here we adopt the approach followed by the authors of \cite{nzt13GA} where they could accommodate all the neutrino parameters within a $A_4$ model by including additional flavons apart from those appearing in usual TBM analysis.  Without assuming any specific realization of $A_4$ symmetry at leading order, we construct the most general neutrino mass matrix in terms of flavons assuming a type I seesaw framework. We then compare this mass matrix with the neutrino mass matrix constructed from the best fit values of neutrino parameters. To reduce the number of free parameters, we assume the $A_4$ triplet flavons to acquire vacuum expectation values (vev) in a way which preserve $Z_2$ or $Z_3$ subgroups of $A_4$. This allows us not only to calculate the $A_4$ flavon vev's in terms of low energy neutrino parameters, but also to calculate the unknown neutrino parameters at low energy. These unknown low energy neutrino parameters include the lightest neutrino mass, which remains undetermined at neutrino experiments. This can be anywhere between 0 and the upper bound set by cosmological bound on the sum of absolute neutrino masses $\sum_i \lvert m_i \rvert < 0.23$ eV \cite{Planck13}. The other unknown neutrino parameters are the leptonic CP violating phases: one Dirac CP phase and two Majorana CP phases. We determine these unknown neutrino parameters as well as the $A_4$ parameters for a particular $A_4$ vacuum alignment using the $3\sigma$ values of experimentally measured neutrino parameters. This is the major advantage as well as motivation for choosing this particular approach as it allows us to determine all the neutrino parameters completely.

After calculating the $A_4$ as well as unknown neutrino parameters for a particular $A_4$ vacuum alignment, we then use the same set of parameters to calculate the baryon asymmetry of the Universe within the framework of leptogenesis. Apart from the origin of fermion mass hierarchy and mixing, the SM also fails to explain the observed baryon asymmetry. Although the BSM framework explaining the baryon asymmetry could be completely decoupled from the one explaining leptonic mass and mixing, it is more economical and predictive if the same model can account for both the observed phenomena. Leptogenesis is one such mechanism which generates the observed baryon asymmetry by creating a leptonic asymmetry first and then converting it into baryon asymmetry through $B+L$ violating electroweak sphaleron transitions \cite{sphaleron}. According to the original proposal of Fukugita and Yanagida \cite{fukuyana}, this mechanism can satisfy all the Sakharov's conditions \cite{sakharov} required to be fulfilled in order to produce a net baryon asymmetry. The out of equilibrium CP violating decay of heavy right handed neutrinos present in the type I seesaw mechanism can naturally produce the required lepton asymmetry. For a review of leptogenesis, one can refer to the review article\cite{davidsonPR}. Some interesting implementation of this idea within several BSM frameworks can be found in \cite{leptoreview}. After calculating the baryon asymmetry in our model with type I seesaw and $A_4$ symmetry, we constrain the model parameters as well as different possible $A_4$ vacuum alignments by comparing our prediction with the observed baryon asymmetry given by \cite{Planck13} \begin{equation}
Y_B = (8.58 \pm 0.22) \times 10^{-11}
\label{barasym}
\end{equation} 
We find that certain combinations of $A_4$ vacuum alignments and light neutrino mass hierarchy (normal or inverted) can not give rise to the observed baryon asymmetry. A few models can give rise to the observed baryon asymmetry for specific values of the lightest neutrino mass and the leptonic CP phases.

To allow experimental verification of the models, we also calculate, for each flavon alignment, the effective neutrino mass $m_{ee} = \lvert \sum_i U^2_{ei} m_i \rvert$ (where $U$ is the leptonic mixing matrix and $m_i$ are light neutrino masses) which can be probed in neutrinoless double beta decay experiments. We find that certain regions of parameter space can be ruled out by phase II of GERDA \cite{GERDA} experiment in future.

This paper is organized as follows. In section \ref{sec:seesaw}, we discuss our model of type I seesaw with $A_4$ flavor symmetry.  In section \ref{sec:lepto}, we briefly discuss the mechanism of leptogenesis. In section \ref{sec:numeric} we describe the numerical analysis adopted here and finally conclude in section \ref{sec:conclude}.

\section{Type I Seesaw with $A_4$ Flavor Symmetry}
\label{sec:seesaw}
Type I seesaw \cite{ti} is the simplest possible realization of the seesaw mechanism where the SM field content is extended by three right handed neutrinos $(\nu^i_R, i=1,2,3)$ singlet under the SM gauge symmetry.  This introduces additional terms in the Yukawa Lagrangian
\begin{eqnarray}
\mathcal{L}_Y = y^{ij}_\nu \ell_i \tilde{H} \nu^j_R +\frac{1}{2} \nu^{iT}_R C^{-1} M^{ij}_R \nu^j_R  +\text{h.c.}
\end{eqnarray}
where $\ell_L \equiv(\nu,~e)^T_L$, $H \equiv (h^0 ,h^-)^T$ and C is the charge conjugation operator. The resulting in $6\times 6$ neutrino mass matrix after electroweak symmetry breaking is given by
\begin{equation}
\mathcal{M}_\nu= \left( \begin{array}{cc}
              0 & M_D   \\
              M^T_D & M_{RR}
                      \end{array} \right) \, ,
\label{eqn:numatrix}       
\end{equation}
where $M_{D}=y_\nu\,v$ is the Dirac neutrino mass and $v$ is the vev of the neutral component of SM higgs doublet. Assuming $M_{RR} \gg M_{D}$, the light neutrino mass is given by the type I seesaw formula 
\begin{equation}
M_{\nu}^I=-M_D M_{RR}^{-1} M_D^{T}
\label{type1eq}
\end{equation}
Assuming the Dirac mass term to be at electroweak scale, one needs $M_{RR}$ to be as heavy as $10^{14}$ GeV in order to generate light neutrino mass of order $0.1$ eV.  The scale of right handed mass can be lowered by suitable fine tuning of the Dirac Yukawa couplings.

We now briefly discuss the $A_4$ realization of type I seesaw mechanism that was presented in the work \cite{nzt13GA}. $A_4$, the group of even permutations of four objects, is also the symmetry group of a tetrahedron. This group has $12$ elements and four irreducible representations with dimensions $n_i$ such that $\sum_i n_i^2=12$. The characters of $4$ representations are shown in table \ref{table:character}. The complex number $\omega$ 
is the cube root of unity. These four representations are denoted by $\bf{1}, \bf{1'}, \bf{1''}$ and $\bf{3}$ respectively. 
\begin{table}[ht]
\centering
\caption{Character table of $A_4$}
\vspace{0.5cm}
\begin{tabular}{ccccc}
 \hline
   Class & $\chi_{1}$ &  $\chi_{1'}$ &  $\chi_{1''}$&  $\chi_{3}$\\ \hline \hline
   $C_1$ & 1&1&1&3 \\ 
    $C_2$& 1&$\omega$&$\omega^2$&0 \\  
$C_3$ &1&$\omega^2$&$\omega$&0\\ 
$C_4$ &1&1&1&-1\\  \hline
\end{tabular}
\label{table:character}
\end{table}

Apart from the SM fields and three right handed neutrino fields required for type I seesaw, the model we are studying has five flavon fields, singlets under SM gauge symmetry, required to break the $A_4$ symmetry as well as to generate the desired structure of lepton mass matrices. The model also has an additional $Z_2$ symmetry to make sure the presence of only the desired terms in the Lagrangian. The transformations of all the fields in the model under $A_4 \times Z_2$ symmetry are shown in table \ref{tab:A4charge}. The $SU(2)_L$ lepton doublets $\ell = (\ell_e, \ell_{\mu}, \ell_{\tau})$ are assumed to transform as triplet $\bf{3}$ under $A_4$ whereas the $SU(2)_L$ singlet charged leptons $e^c, \mu^c, \tau^c$ transform as $\bf{1}, \bf{1''}, \bf{1'}$ respectively. The $SU(2)_L$ singlet right handed neutrinos $\nu_R$ transform as a triplet under $A_4$. The SM higgs field $H$ transforms like a singlet $\bf{1}$ under $A_4$ whereas the SM singlet flavon fields $\phi_E, \phi_N, \eta, \chi, \psi$ transform as $\bf{3}, \bf{3}, \bf{1}, \bf{1'}, \bf{1''}$ respectively. 

\begin{table}[h]
\begin{tabular}{|l|l|l|l|l|l|l|l|l|l|l|l|}
\hline
&$H$  &$l$ &$\nu_R$ & $e_R$  & $\mu_R$ & $\tau_R$ &$\phi_E$  & $ \phi_N$  &$ \eta$  &$\chi$  &$\psi$    \\ \hline
 $A_4$& 1 & 3 & 3 & 1 & $1^{\prime\prime}$  &$1^{\prime}$  & 3 &  3& 1 & $1^{\prime}$ & $1^{\prime\prime}$ \\ 
$Z_2$ & 1 & 1 & 1 & -1 &-1  & -1 & -1 & 1 & 1 & 1 &1  \\ \hline
\end{tabular}
\caption{Transformation of the fields under $A_4 \times Z_2$ symmetry of the model}
\label{tab:A4charge}
\end{table}

After fixing the transformation of the model fields under the flavor symmetry as well as the SM gauge symmetry, the Lagrangian for the lepton can be written as
\begin{widetext}
\begin{equation}
\begin{split}
 L \supset \left( H\bar{\ell}(\lambda_ee_R+\lambda_\mu\mu_R+\lambda_\tau\tau_R)(\frac{\phi_E}{\Lambda}) +\lambda_N\tilde{H}\bar{\ell}\nu_R+\text{h.c.} \right)\\ +\Lambda_{RR}\nu^T_R \nu_R\left( \frac{c_N\phi_N+c_\eta\eta+c_\psi\psi+c_\chi\chi}{\Lambda} \right)+\text{h.c.}
 \label{YukawaL}
\end{split}
 \end{equation}
 \end{widetext}
where $\Lambda$ is the scale at which the flavons acquire vev's in order to break the $A_4$ symmetry. The dimensionless couplings $c_N,c_\eta,c_\chi$ and $c_\psi$ are, in general, complex. Due to the non-trivial $Z_2$ charge assignments, the flavon field $\phi_E$ couples only to the charged leptons. We can decompose the terms in the Lagrangian above into $A_4$ singlets using the $A_4$ product rules given in appendix \ref{appen1}. Similar to the leptonic Lagrangian, one should also decompose the full scalar potential of the model into $A_4$ singlets and find out the vacuum alignments of the flavon fields by minimizing the potential. Here we do not perform this detailed exercise which can be found elsewhere. Instead, we assume specific vacuum alignments of the flavon fields and study their phenomenological consequences. Assuming the triplet flavon $\phi_E$ to acquire vev as $\langle \phi_E \rangle = \Lambda (1,0,0)^T$, we can write down the charged lepton mass matrix in the diagonal form as
\begin{equation}
M_{\ell}=\left(\begin{array}{ccc}
\lambda_e& 0&0\\
0& \lambda_{\mu} & 0 \\
0 & 0 & \lambda_{\tau}
\end{array}\right)v
\label{mclepton}
\end{equation}
where $v$ is the vev of the SM higgs doublet as mentioned earlier. Similarly, decomposing the term $\lambda_N \bar{H}\bar{\ell}\nu_R$ into $A_4$ singlets, the Dirac neutrino mass matrix can be written as
\begin{equation}
M_D=\left(\begin{array}{ccc}
\lambda_Nv& 0&0\\
0& 0 & \lambda_{N}v \\
0 & \lambda_{N}v& 0
\end{array}\right)
\label{mdirac}
\end{equation}
The right handed neutrino mass matrix can receive contribution from four different flavons $\phi_N, \eta, \chi, \psi$ as seen from the leptonic Lagrangian \eqref{YukawaL}. Adopting the same notations as used by the authors of \cite{nzt13GA}, we denote the triplet flavon vev's as $c_N \langle \phi_N \rangle = \Lambda (\phi_a, \phi_b, \phi_c)$ and the singlet flavon vev's as $c_\eta \langle \eta \rangle = \Lambda \eta, c_\chi \langle \chi \rangle = \Lambda \chi, c_\psi \langle \psi \rangle = \Lambda \psi$. Thus, in this notation $\phi_{a,b,c}, \eta, \chi, \psi$ are dimensionless and have absorbed the respective Yukawa couplings. In this notation, the right handed neutrino mass matrix can be written as
\begin{equation}
M_{RR}=\Lambda_{RR}\left(\begin{array}{ccc}
\frac{2}{3}\phi_a+\eta&-\frac{1}{3}\phi_c+\psi&-\frac{1}{3}\phi_b+\chi\\
-\frac{1}{3}\phi_c+\psi&\frac{2}{3}\phi_b+\chi & -\frac{1}{3}\phi_a+\eta \\
-\frac{1}{3}\phi_b+\chi &-\frac{1}{3}\phi_a+\eta&\frac{2}{3}\phi_c+\psi
\end{array}\right)
\label{mright}
\end{equation}
We can use the Dirac and right handed Majorana mass matrices given by equations \eqref{mdirac} and \eqref{mright} respectively in the type I seesaw formula given by equation \eqref{type1eq} to construct the light neutrino mass matrix in terms of $A_4$ parameters. In the section on numerical analysis \ref{sec:numeric}, we will discuss how these $A_4$ parameters can be computed in terms of the light neutrino parameters by comparing this mass matrix with the one constructed using light neutrino data.

\section{Leptogenesis}
\label{sec:lepto}
The observed Universe at present is baryon asymmetric, with the ratio of measured excess of baryons over anti-baryons to the entropy density is given by equation \eqref{barasym}.  If the Universe had started in a baryon symmetric way, three conditions must be satisfied in order to create a net baryon asymmetry. As pointed out first by Sakharov \cite{sakharov}, these three conditions are (i) Baryon number violation, (ii) C and CP violation and (iii) Departure from equilibrium. Although the standard model satisfies these conditions in an expanding Universe like ours, the amount of CP violation measured in the SM quark sector turns out to be too small to account for the entire baryon asymmetry of the Universe. This extra source of CP violation could be the leptonic sector which is not yet experimentally determined. Leptogenesis provides a minimal setup to connect lepton sector CP violation with the observed baryon asymmetry and also put indirect limits on these CP phases from the requirement of producing correct baryon asymmetry. 

\begin{figure}[ht]
\centering
\includegraphics[width=0.8\textwidth]{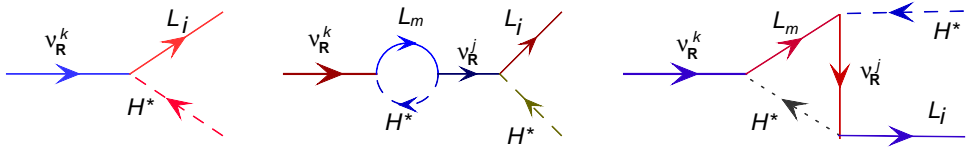}
\caption{Right handed neutrino decay in type I seesaw models}
\label{lepto1}
\end{figure}

In a model with type I seesaw mechanism of neutrino masses, the CP violating out of equilibrium decay of the heavy right handed neutrinos into SM particles (as shown in figure \ref{lepto1}) can give rise to an asymmetry in the leptonic sector. This asymmetry is given by  
\begin{equation}
\epsilon_{\nu^k_R} = \sum_i \frac{\Gamma(\nu^k_R \rightarrow L_i +H^*)-\Gamma (\nu^k_R \rightarrow \bar{L_i}+H)}{\Gamma(\nu^k_R \rightarrow L_i +H^*)
+\Gamma (\nu^k_R \rightarrow \bar{L_i}+H)}
\end{equation}
If the three right handed neutrinos are assumed to be hierarchical such that $M_{2,3} \gg M_1$, it is sufficient to consider the asymmetry produced by the decay of the lightest right handed neutrino $\nu^1_R$. Adopting the notations of \cite{joshipura}, the lepton asymmetry resulting from the decay of $\nu^1_R$ can be written as
\begin{eqnarray}
\epsilon^{\alpha}_1 &=& \frac{1}{8\pi v^2}\frac{1}{(M^{\dagger}_{D}M_{D})_{11}} \sum_{j=2,3} \text{Im}[(M^*_{D})_{\alpha 1}
(M^{\dagger}_{D}M_{D})_{1j}(M_{D})_{\alpha j}]g(x_j) \nonumber \\&& + \frac{1}{8\pi v^2}\frac{1}{(M^{\dagger}_{D}M_{D})_{11}} 
\sum_{j=2,3} \text{Im}[(M^*_{D})_{\alpha 1}(M^{\dagger}_{D}M_{D})_{j1}(M_{D})_{\alpha j}]\frac{1}{1-x_j}
\label{eps1}
\end{eqnarray}
where $v = 174 \; \text{GeV}$ is the vev of the Higgs doublet responsible for breaking the electroweak symmetry, $$ g(x) = \sqrt{x} 
\left (1+\frac{1}{1-x}-(1+x)\text{ln}\frac{1+x}{x} \right) $$and $x_j = M^2_j/M^2_1$. The second term in the expression for $\epsilon^{\alpha}_1$ 
above vanishes when summed over all the flavors $\alpha = e, \mu, \tau$. The sum over flavors is given by
\begin{equation}
\epsilon_1 = \frac{1}{8\pi v^2}\frac{1}{(M^{\dagger}_{D}M_{D})_{11}}\sum_{j=2,3} \text{Im}[(M^{\dagger}_{D}M_{D})^2_{1j}]g(x_j)
\label{noflavor}
\end{equation}

The above leptonic asymmetry can be converted into baryon asymmetry as 
\begin{equation}
Y_B = c \kappa \frac{\epsilon_1}{g_*}
\end{equation}
through electroweak sphaleron processes \cite{sphaleron}. Here, $c$ is a measure of the fraction of lepton asymmetry being 
converted into baryon asymmetry and is approximately equal to $-0.55$. $\kappa$ is the dilution factor due to wash-out processes which 
erase the produced asymmetry and can be parametrized as \cite{kolbturner}
\begin{eqnarray}
-\kappa &\simeq &  \sqrt{0.1K} \text{exp}[-4/(3(0.1K)^{0.25})], \;\; \text{for} \; K  \ge 10^6 \nonumber \\
&\simeq & \frac{0.3}{K (\ln K)^{0.6}}, \;\; \text{for} \; 10 \le K \le 10^6 \nonumber \\
&\simeq & \frac{1}{2\sqrt{K^2+9}},  \;\; \text{for} \; 0 \le K \le 10.
\end{eqnarray}
where K is given as
$$ K = \frac{\Gamma_1}{H(T=M_1)} = \frac{(M^{\dagger}_{D}M_{D})_{11}M_1}{8\pi v^2} \frac{M_{Pl}}{1.66 \sqrt{g_*}M^2_1} $$
Here $\Gamma_1$ is the decay width of $N_1$ and $H(T=M_1)$ is the Hubble constant at temperature $T = M_1$. $M_{Pl}$ is the Planck mass and the factor $g_*$ is the 
effective number of relativistic degrees of freedom at $T=M_1$ which is approximately $110$.

The lepton asymmetry given by equation (\ref{noflavor}) is obtained by summing over all lepton flavors $\alpha = e, \mu, \tau$.  At very high temperatures $(T \geq 10^{12} \text{GeV})$ all charged 
lepton flavors are out of equilibrium and hence all of them behave similarly resulting in the one flavor regime of leptogenesis discussed above. However, at temperatures 
$ T < 10^{12}$ GeV $(T < 10^9 \text{GeV})$, interactions involving tau (muon) Yukawa couplings enter equilibrium and flavor effects become 
important as discussed in details by the authors of \cite{flavorlepto}. Taking these flavor effects into account, the final baryon asymmetry can be written as 
\begin{equation}
Y^{2 flavor}_B = \frac{-12}{37g^*}[\epsilon_2 \eta\left (\frac{417}{589}\tilde{m_2} \right)+\epsilon^{\tau}_1\eta\left (\frac{390}{589}
\tilde{m_{\tau}}\right )] \nonumber
\end{equation}
\begin{equation}
Y^{3 flavor}_B = \frac{-12}{37g^*}[\epsilon^e_1 \eta\left (\frac{151}{179}\tilde{m_e}\right)+ \epsilon^{\mu}_1 \eta\left (\frac{344}{537}
\tilde{m_{\mu}}\right)+\epsilon^{\tau}_1\eta\left (\frac{344}{537}\tilde{m_{\tau}} \right )] \nonumber
\end{equation}
where $\epsilon_2 = \epsilon^e_1 + \epsilon^{\mu}_1, \tilde{m_2} = \tilde{m_e}+\tilde{m_{\mu}}, \tilde{m_{\alpha}} = \frac{(M^*_{D})_{\alpha 1} 
(M_{D})_{\alpha 1}}{M_1}$. The function $\eta$ is given by 
$$ \eta (\tilde{m_{\alpha}}) = \left [\left ( \frac{\tilde{m_{\alpha}}}{8.25 \times 10^{-3} \text{eV}} \right )^{-1}+ \left ( \frac{0.2\times 
10^{-3} \text{eV}}{\tilde{m_{\alpha}}} \right )^{-1.16} \right ]^{-1} $$

For the calculation of baryon asymmetry, we first calculate the right handed neutrino mass spectrum by diagonalizing the right handed singlet neutrino mass matrix $M_{RR}$ as
\begin{equation}
U^*_R M_{RR} U^{\dagger}_R = \text{diag}(M_1, M_2, M_3)
\label{mrrdiag}
\end{equation}
In this diagonal $M_{RR}$ basis, according to the type I seesaw formula, the Dirac neutrino mass matrix also changes to 
\begin{equation}
m_{LR} = m^0_{LR} U_R
\label{mlrdiag}
\end{equation}
where $m^0_{LR}$ is the Dirac neutrino mass matrix given by equation \eqref{mdirac} in this model.

\section{Numerical Analysis}
\label{sec:numeric}
The light neutrino mass matrix can be constructed using the neutrino data of mixing angles and mass squared differences. The Pontecorvo-Maki-Nakagawa-Sakata (PMNS) leptonic mixing matrix is related to the diagonalizing 
matrices of neutrino and charged lepton mass matrices $U_{\nu}, U_{\ell}$ respectively, as
\begin{equation}
U_{\text{PMNS}} = U^{\dagger}_{\ell} U_{\nu}
\label{pmns0}
\end{equation}
The PMNS mixing matrix can be parametrized as
\begin{equation}
U_{\text{PMNS}}=\left(\begin{array}{ccc}
c_{12}c_{13}& s_{12}c_{13}& s_{13}e^{-i\delta}\\
-s_{12}c_{23}-c_{12}s_{23}s_{13}e^{i\delta}& c_{12}c_{23}-s_{12}s_{23}s_{13}e^{i\delta} & s_{23}c_{13} \\
s_{12}s_{23}-c_{12}c_{23}s_{13}e^{i\delta} & -c_{12}s_{23}-s_{12}c_{23}s_{13}e^{i\delta}& c_{23}c_{13}
\end{array}\right) U_{\text{Maj}}
\label{matrixPMNS}
\end{equation}
where $c_{ij} = \cos{\theta_{ij}}, \; s_{ij} = \sin{\theta_{ij}}$ and $\delta$ is the Dirac CP phase. The diagonal matrix $U_{\text{Maj}}=\text{diag}(1, e^{i\alpha}, e^{i(\beta+\delta)})$  contains the Majorana CP phases $\alpha, \beta$. In the model we are working, the charged lepton mass matrix is diagonal given by equation \eqref{mclepton} and hence $U_{\ell} = I$, the identity matrix. Therefore, in the diagonal charged lepton basis $U_{\text{PMNS}} = U_{\nu} $. The light neutrino mass matrix can now be written as 
\begin{equation}
M_\nu=U_{\text{PMNS}}M^{\text{diag}}_{\nu}U^T_{\text{PMNS}}
\label{mnu1}
\end{equation}
where $M^{\text{diag}}_{\nu} = \text{diag}(m_1, m_2, m_3)$ is the diagonal form of light neutrino mass matrix. For the case of normal hierarchy, the three neutrino mass eigenvalues can be written as $M^{\text{diag}}_{\nu} 
= \text{diag}(m_1, \sqrt{m^2_1+\Delta m_{21}^2}, \sqrt{m_1^2+\Delta m_{31}^2})$, while for the case of inverted hierarchy (IH), it can be written as 
$M^{\text{diag}}_{\nu} = \text{diag}(\sqrt{m_3^2+\Delta m_{23}^2-\Delta m_{21}^2}, \sqrt{m_3^2+\Delta m_{23}^2}, m_3)$.
If the light neutrino mass is given by the type I seesaw formula \eqref{type1eq}, then the right handed neutrino mass matrix can be obtained by inverting the type I seesaw formula
\begin{equation}
M_{RR}=M^T_DU_{\text{PMNS}}(M^{diag}_\nu)^{-1}U^T_{\text{PMNS}}M_D
\label{mright2}
\end{equation}
Thus we have two expressions for $M_{RR}$, one completely in terms of $A_4$ parameters given by equation \eqref{mright} and the other in terms of light neutrino parameters and Dirac neutrino mass \eqref{mright2}. The right handed Majorana mass matrix $M_{RR}$ is complex symmetric and hence six independent complex parameters. Thus, comparing $M_{RR}$ given by equations \eqref{mright} and \eqref{mright2}, we can write down six independent equations relating $A_4$ parameters with the light neutrino ones. Solving these equations give us the six $A_4$ flavon parameters $\phi_a, \phi_b, \phi_c, \eta, \chi, \psi$ given in Appendix \ref{appen2}. It should be noted that, in the earlier work \cite{nzt13GA}, hermitian conjugate of the diagonalizing matrices were used instead of transpose in the definitions of light and heavy neutrino mass matrices given in equation (\ref{mnu1}) and equation (\ref{mright2}). Since the Majorana mass matrices are complex symmetric instead of hermitian, using transpose of the diagonalizing matrices is more appropriate and hence we have adopted that convention. This change will also give rise to changes in the equations relating $A_4$ and neutrino parameters shown in Appendix \ref{appen2} from the ones given in the earlier work \cite{nzt13GA}. This will also change the solutions of the equations relating different flavon vev's for specific $A_4$ triplet vacuum alignments, which we discuss below. 

The light neutrino mass matrix constructed from the neutrino oscillation parameters has four free parameters namely, the lightest neutrino mass $m_{\text{lightest}}$, one Dirac CP phase $\delta$ and two Majorana CP phases $\alpha, \beta$ which remain undetermined experimentally till now. If we fix these four parameters, then the six $A_4$ parameters can be determined using the equations given in Appendix \ref{appen2} upto a factor $F=\frac{v^2\lambda^2_{N}}{\Lambda_{RR}}$. We can reduce the number of these free parameters in the light neutrino sector, if we consider some specific flavon vev alignments like the ones discussed by the authors of \cite{nzt13GA}. These specific alignments of the $A_4$ triplet vev's $(\phi_a, \phi_b, \phi_c)$ can keep the $Z_2$ or $Z_3$ subgroup of $A_4$ unbroken in the flavon space. Similar to the work \cite{nzt13GA}, here also we do our calculations for both $Z_2, Z_3$ preserving and $Z_2, Z_3$ breaking triplet vev patterns. These triplet vev patterns are listed in the table \ref{vevA4}. 
\begin{table}[h]
\begin{center}
\begin{tabular}{|l||l|l|}
\hline
\bf{Subgroup} & \multicolumn{2}{l|}{ \bf{$A_4$ triplet vev alignment} } \\ \hline\hline
$Z_2$ & \multicolumn{2}{l|}{(1,0,0),(0,1,0),(0,0,1)} \\ \hline\hline
$Z_3$ & \multicolumn{2}{l|}{(-1,1,1),(1,1,-1),(1,-1,1),(1,1,1)} \\ \hline\hline
Breaking& \multicolumn{2}{l|}{(0,1,1),(1,0,1),(1,1,0),(0,1,-1),(2,1,1),(1,1,2),(1,2,1),(1,-2,1),(1,1,-2),(-2,1,1)} \\ \hline
\end{tabular}
\end{center}
\caption{$A_4$ triplet vev alignments that preserve as well as break a $Z_2$ or $Z_3$ subgroup of $A_4$}
\label{vevA4}
\end{table}
Each of these vev alignments give rise to two equations involving $\phi_a, \phi_b, \phi_c$. Since $\phi_a, \phi_b, \phi_c$ are complex, these two complex equations give rise to four real equations which can be solved numerically to determine the four light neutrino parameters $m_{\text{lightest}}, \delta, \alpha, \beta$. For example, the vev alignment $(1,0,0)$ gives $\phi_a = 1, \phi_b = 0, \phi_c = 0$ where $\phi_{a,b,c}$ are given by equations \eqref{eq1appen2}, \eqref{eq2appen2}, \eqref{eq3appen2} respectively in appendix \ref{appen2}. After cancelling out the common factor $F$ from the last two equations, we are left with two complex equations involving only the light neutrino parameters which can be found out by solving these equations numerically. In all the cases of vev alignments listed in table \ref{vevA4}, the common factor $F$ in $\phi_{a,b,c}$ can be cancelled out. After finding the four neutrino parameters numerically, we calculate the remaining three singlet $A_4$ flavon vev's $\eta, \chi, \psi$. The common factor $F$ can also be calculated numerically for each of these cases. 

We use the values of $A_4$ parameters determined for a particular triplet vev alignment in the right handed Majorana neutrino mass matrix given by equation \eqref{mright}. If we use the analytical expression for the factor $F = \frac{v^2\lambda^2_{N}}{\Lambda_{RR}}$, then the mass matrix $M_{RR}$ can be calculated upto a numerical factor $\lambda^2_{N}$. It should be noted that although the parameter $F$ gets fixed numerically for each possible triplet vev alignment, we still have the freedom to choose either $\lambda_{N}$ or $\Lambda_{RR}$. We choose $\lambda_N$, the Dirac Yukawa coupling of neutrinos, in such a way that the lightest right handed neutrino mass falls in the appropriate flavor regime of leptogenesis. The chosen values of $\lambda^2_{N}$ are listed in table \ref{lambdaN}. Using these numerical values of the model parameters, we then compute the baryon asymmetry for all the cases of $A_4$ vacuum alignments discussed above.

It should be noted that in the earlier work \cite{nzt13GA}, the most general set of unknown neutrino parameters $(m_{\text{lightest}}, \delta, \alpha, \beta)$ were not calculated. Rather, the authors of \cite{nzt13GA} considered specific choices of Majorana CP phases. Here we compute the most general set of such solutions in agreement with the $A_4$ triplet vacuum alignments.

\begin{center}
\begin{table}[htb]
\begin{tabular}{|c|c|c|c|}
\hline
& 1 Flavor& 2 Flavor & 3 Flavor \\
\hline
$\lambda^2_N$ & $1$ & $1.5\times10^{-3} $ &$1.5\times10^{-5}$\\

\hline
\end{tabular}
\caption{Values of $\lambda^2_N$ for different flavor regime} 
\label{lambdaN}
\end{table}
\end{center}
\begin{figure}[p]
\centering
$
\begin{array}{ccc}

\includegraphics[width=0.9\textwidth]{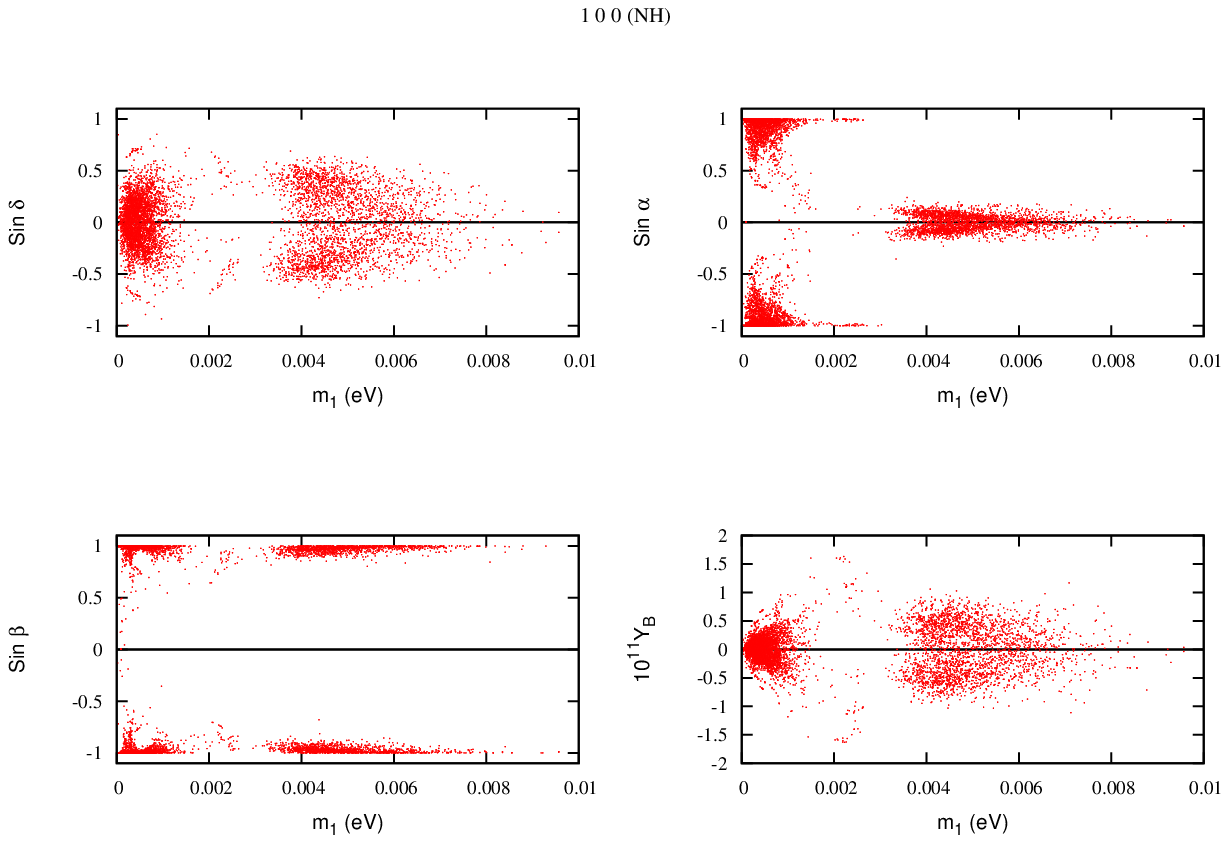} 
 \\
\end{array}$
 \caption{Normal Hierarchy with triplet flavon vev alignment $(1,0,0)$}
 \label{fig1}
\end{figure}
\begin{figure}[p]
\centering
$
\begin{array}{ccc}

\includegraphics[width=0.9\textwidth]{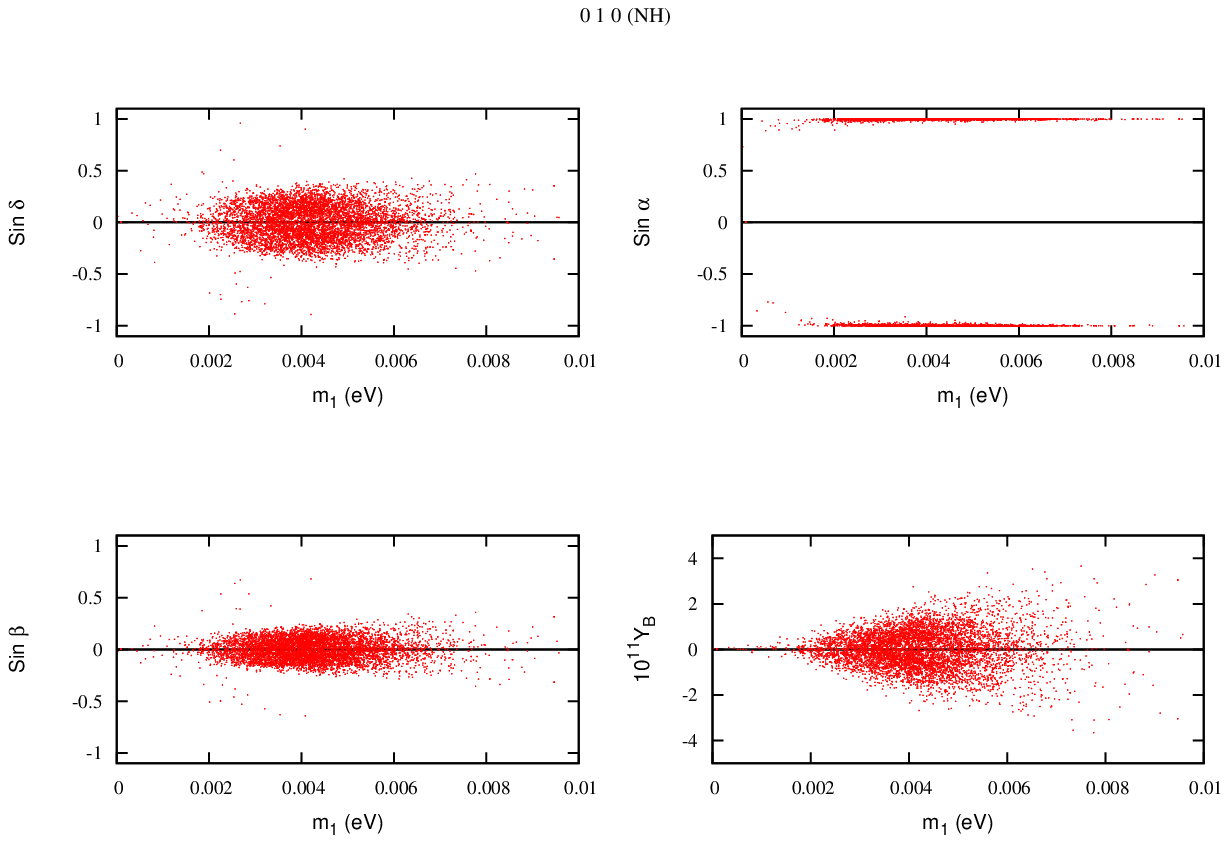} 
 \\
\end{array}$
 \caption{Normal Hierarchy with triplet flavon vev alignment $(0,1,0)$}
  \label{fig2}
\end{figure}
\begin{figure}[p]
\centering
$
\begin{array}{ccc}

\includegraphics[width=0.9\textwidth]{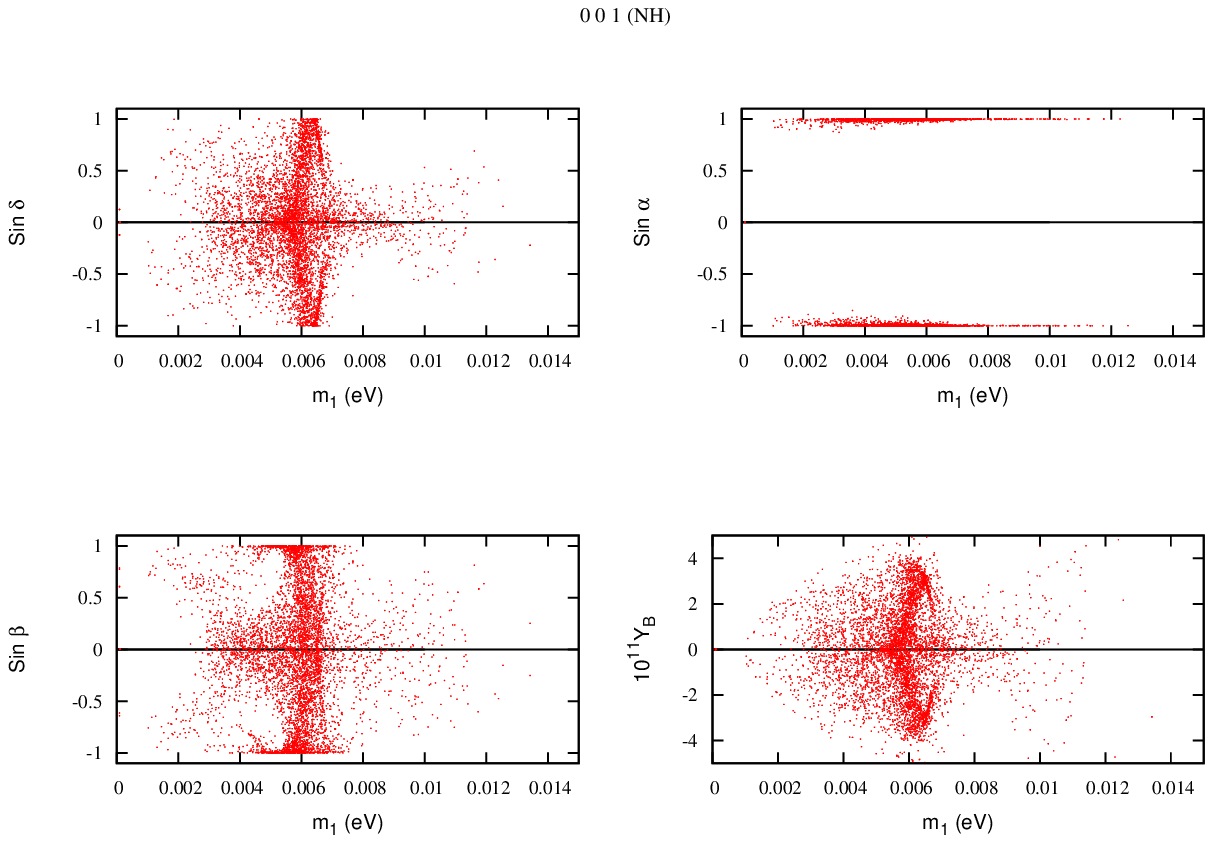} 
 \\
\end{array}$
 \caption{Normal Hierarchy with triplet flavon vev alignment $(0,0,1)$}
  \label{fig3}
\end{figure}
\begin{figure}[p]
\centering
$
\begin{array}{ccc}

\includegraphics[width=0.9\textwidth]{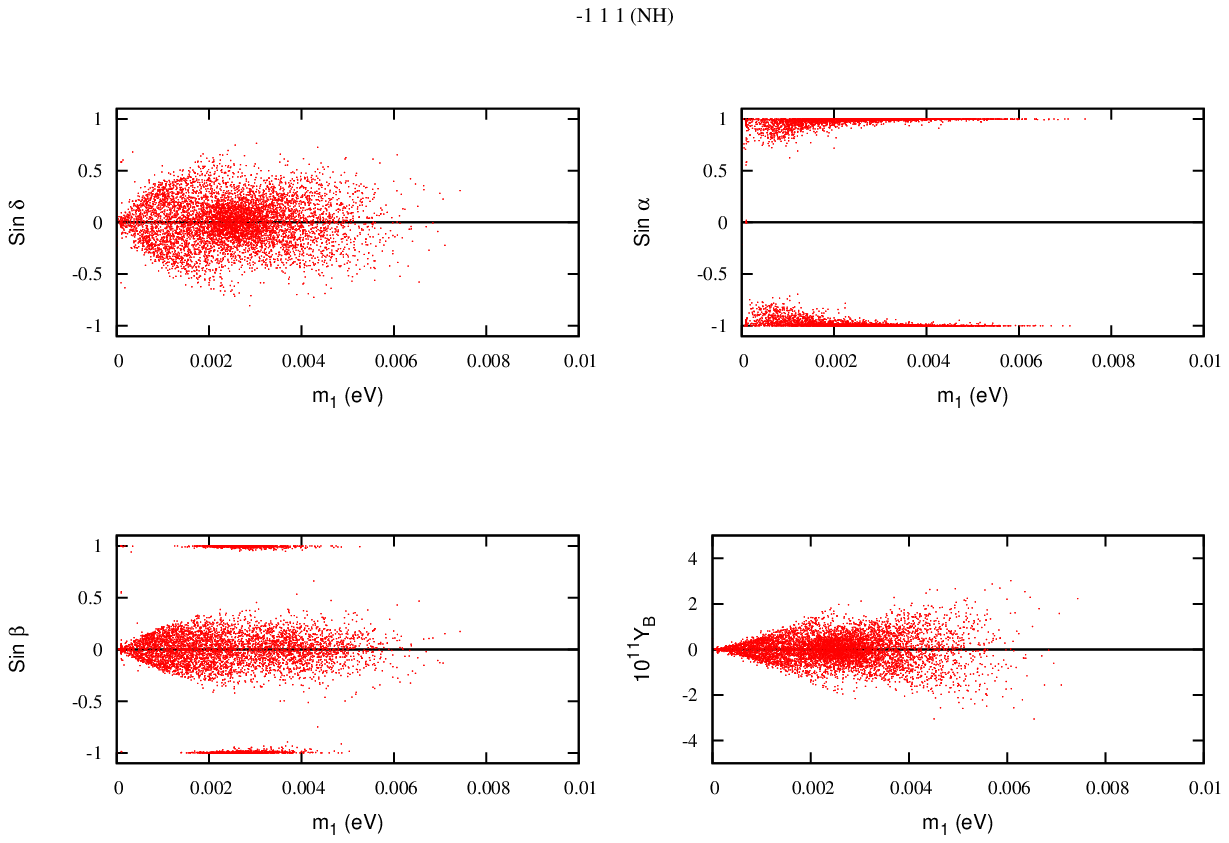} 
 \\
\end{array}$
 \caption{Normal Hierarchy with triplet flavon vev alignment $(-1,1,1)$}
  \label{fig4}
\end{figure}
\begin{figure}[p]
\centering
$
\begin{array}{ccc}

\includegraphics[width=0.9\textwidth]{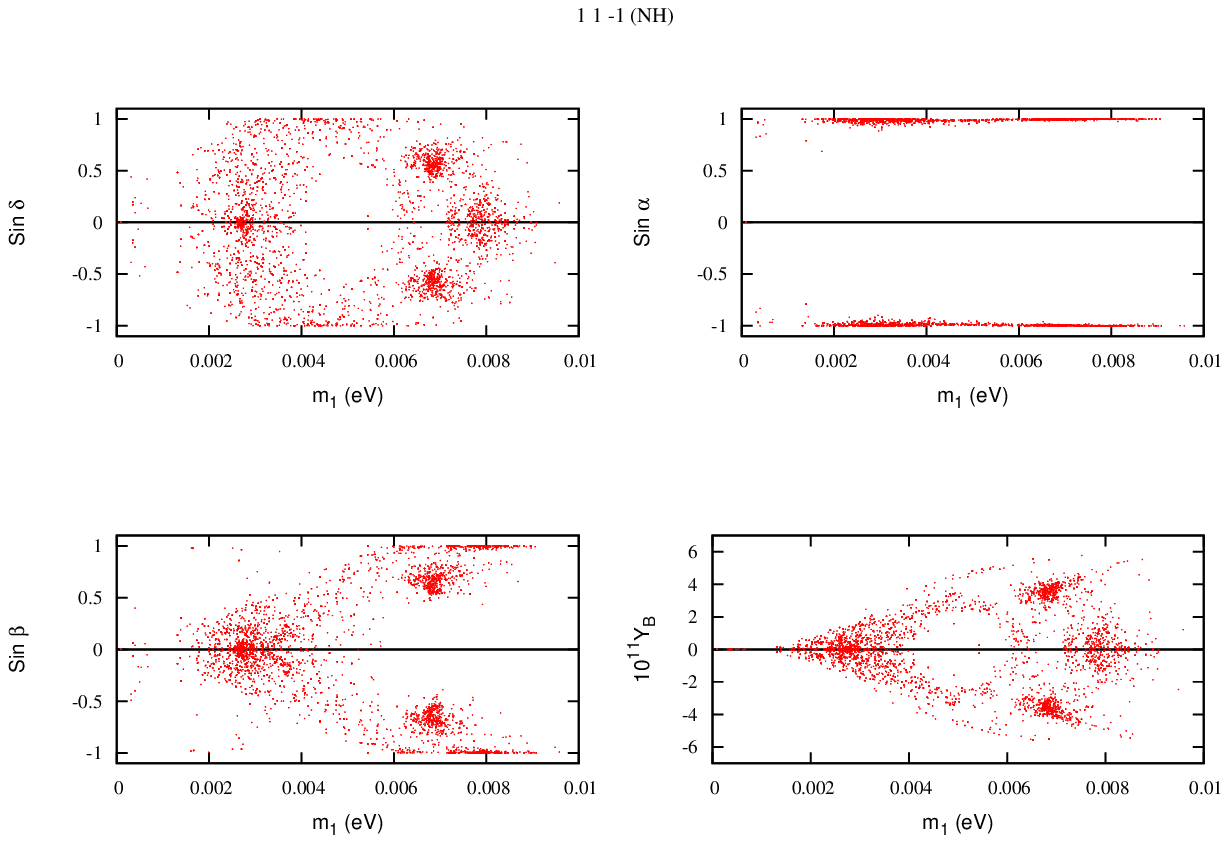} 
 \\
\end{array}$
 \caption{Normal Hierarchy with triplet flavon vev alignment $(1,1,-1)$}
  \label{fig5}
\end{figure}
\begin{figure}[p]
\centering
$
\begin{array}{ccc}

\includegraphics[width=0.9\textwidth]{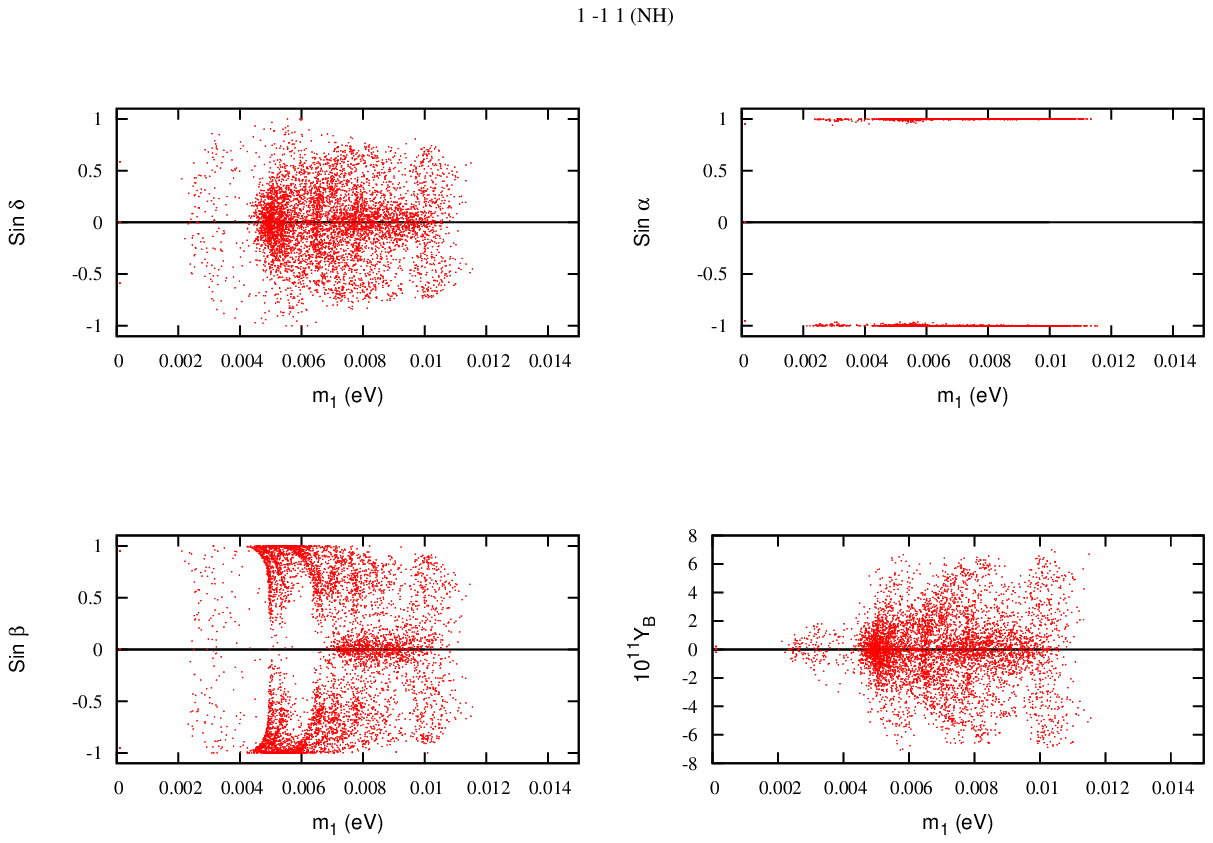} 
 \\
\end{array}$
 \caption{Normal Hierarchy with triplet flavon vev alignment $(1,-1,1)$}
  \label{fig6}
\end{figure}
\begin{figure}[p]
\centering
$
\begin{array}{ccc}

\includegraphics[width=0.9\textwidth]{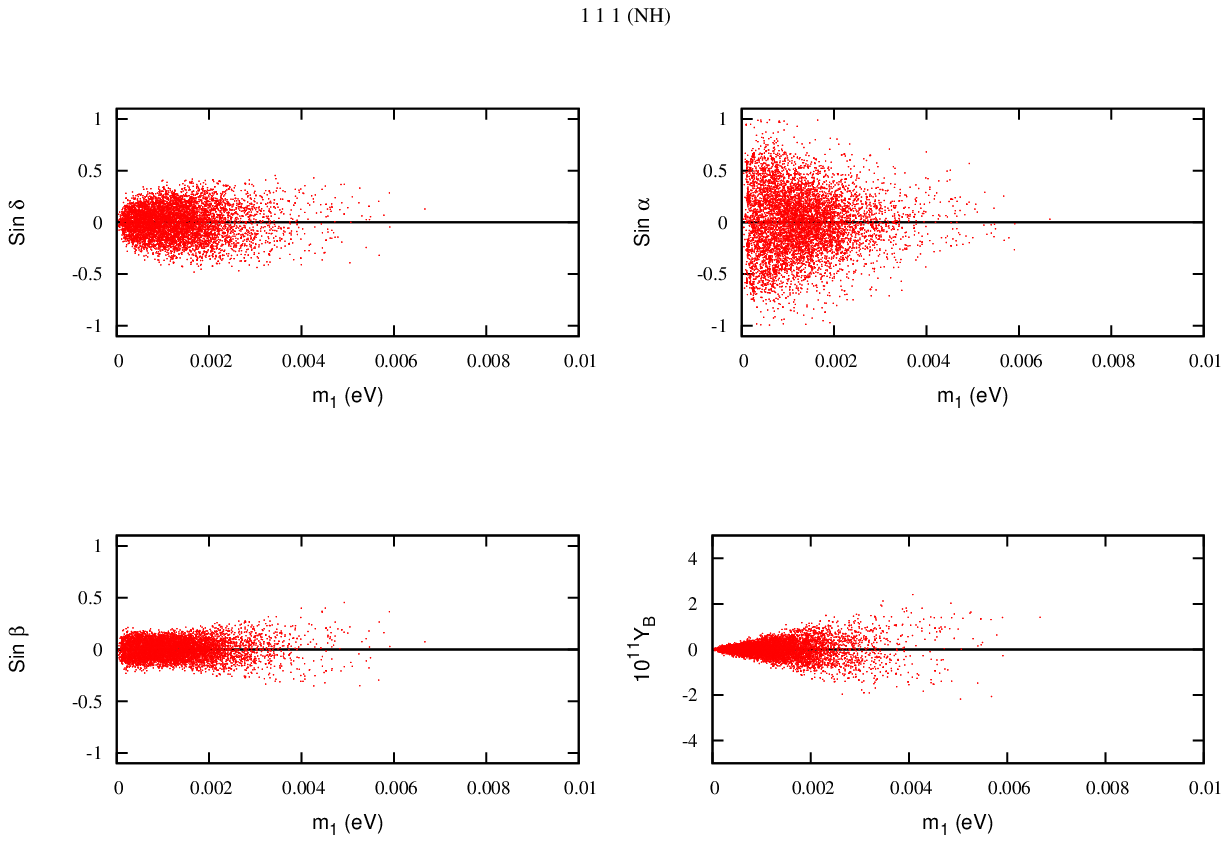} 
 \\
\end{array}$
 \caption{Normal Hierarchy with triplet flavon vev alignment $(1,1,1)$}
  \label{fig7}
\end{figure}

\begin{figure}[p]
\centering
$
\begin{array}{ccc}

\includegraphics[width=0.9\textwidth]{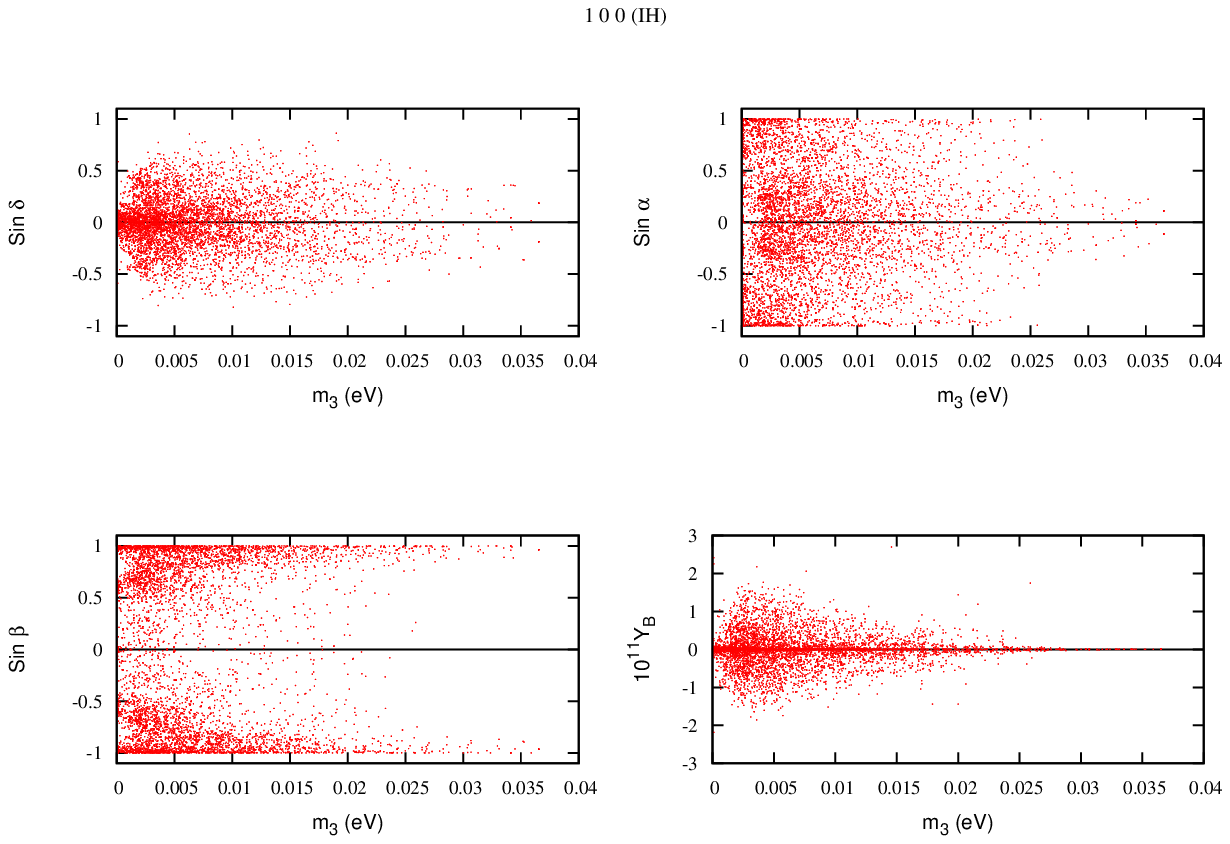} 
 \\
\end{array}$
 \caption{Inverted Hierarchy with triplet flavon vev alignment $(1,0,0)$}
  \label{fig18}
\end{figure}
\begin{figure}[p]
\centering
$
\begin{array}{ccc}

\includegraphics[width=0.9\textwidth]{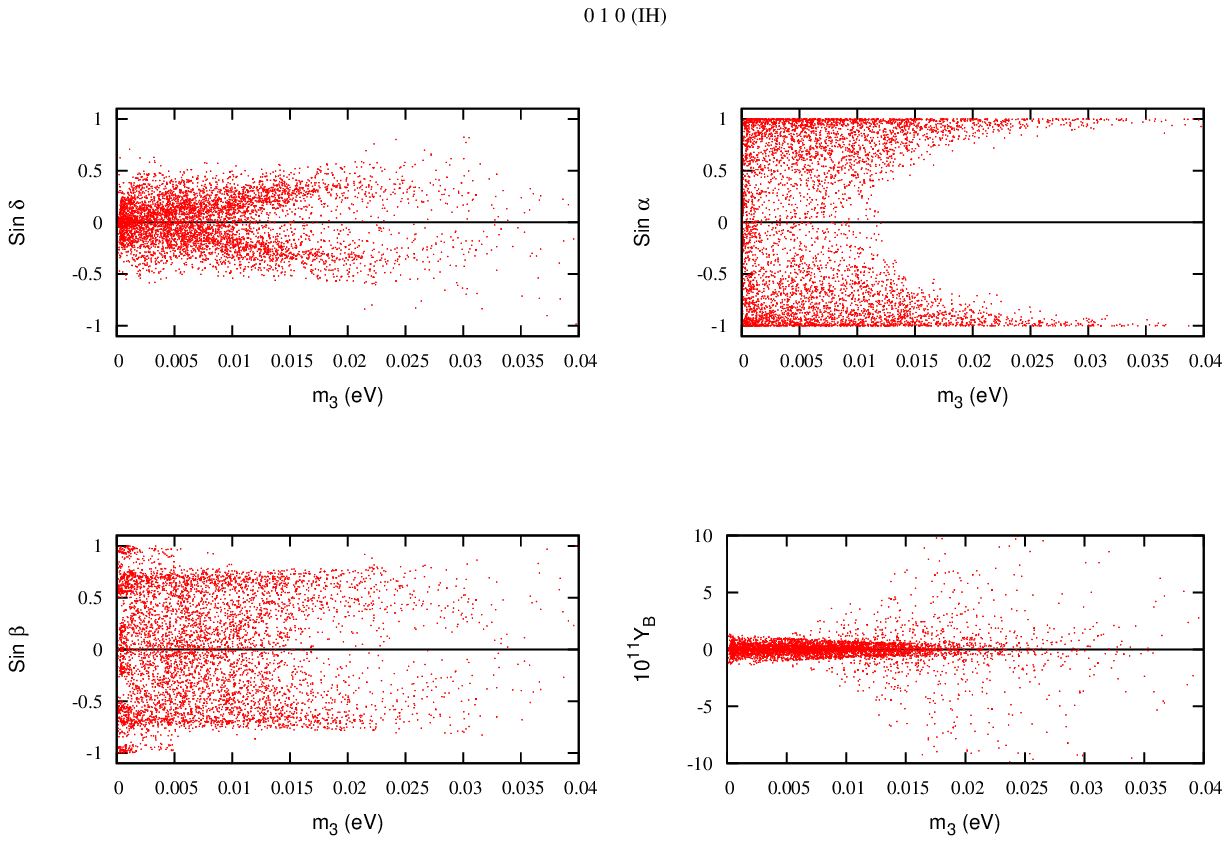} 
 \\
\end{array}$
 \caption{Inverted Hierarchy with triplet flavon vev alignment $(0,1,0)$}
  \label{fig19}
\end{figure}
\begin{figure}[p]
\centering
$
\begin{array}{ccc}

\includegraphics[width=0.9\textwidth]{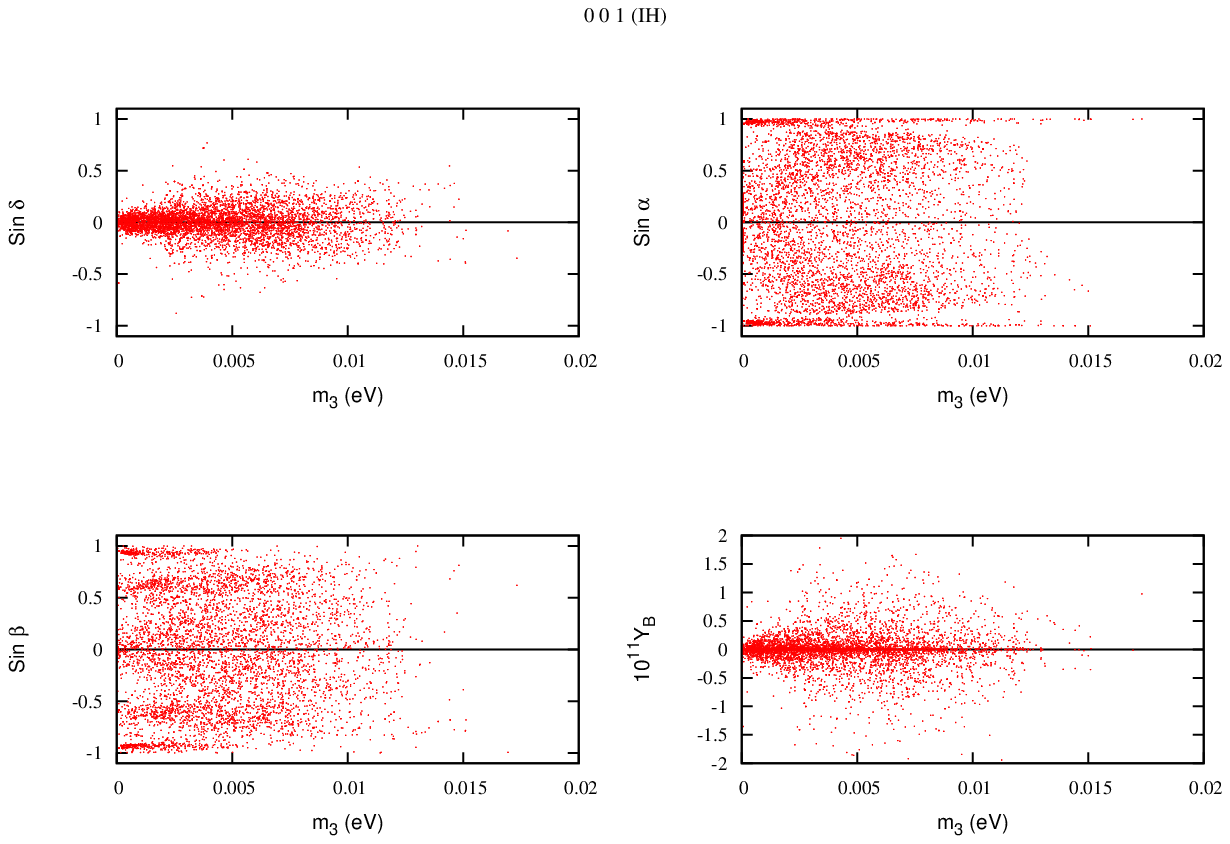} 
 \\
\end{array}$
 \caption{Inverted Hierarchy with triplet flavon vev alignment $(0,0,1)$}
  \label{fig20}
\end{figure}
\begin{figure}[p]
\centering
$
\begin{array}{ccc}

\includegraphics[width=0.9\textwidth]{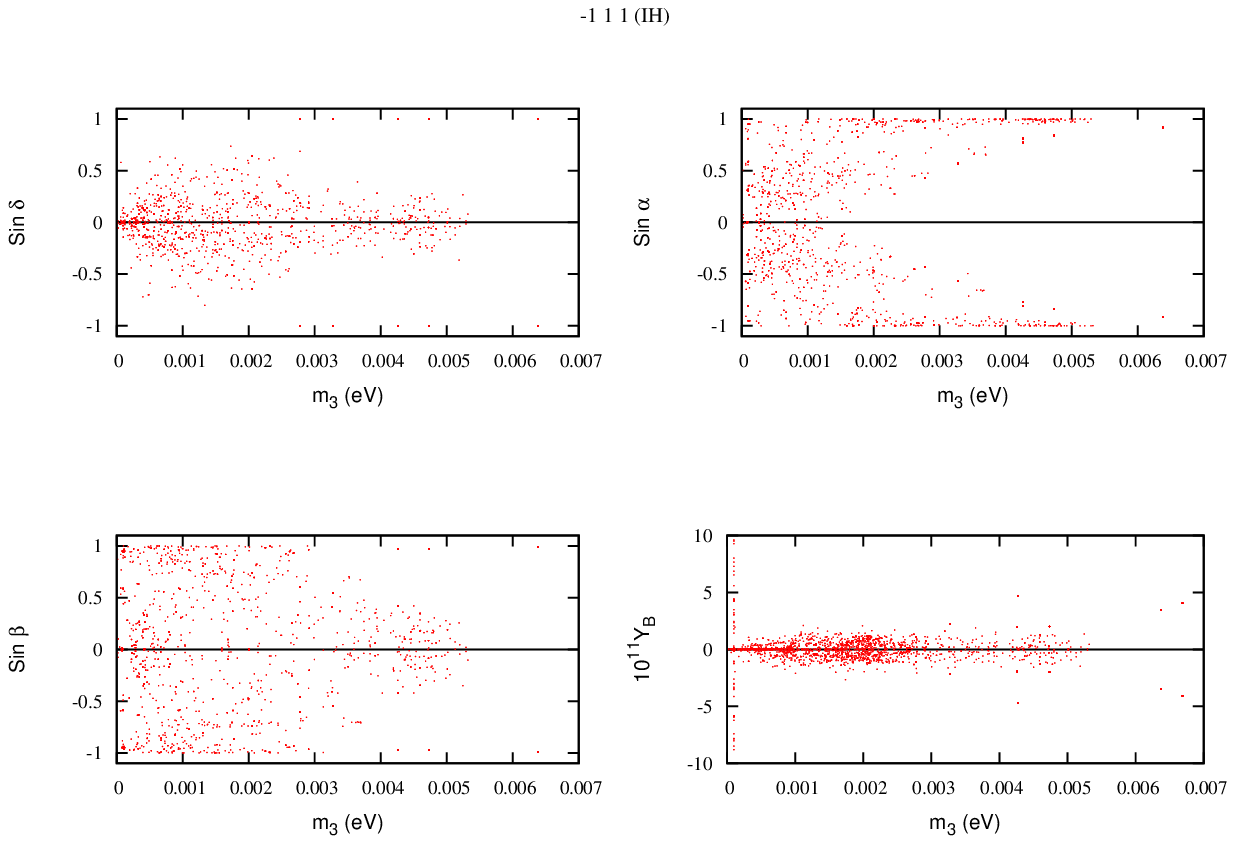} 
 \\
\end{array}$
 \caption{Inverted Hierarchy with triplet flavon vev alignment $(-1,1,1)$}
  \label{fig21}
\end{figure}
\begin{figure}[p]
\centering
$
\begin{array}{ccc}

\includegraphics[width=0.9\textwidth]{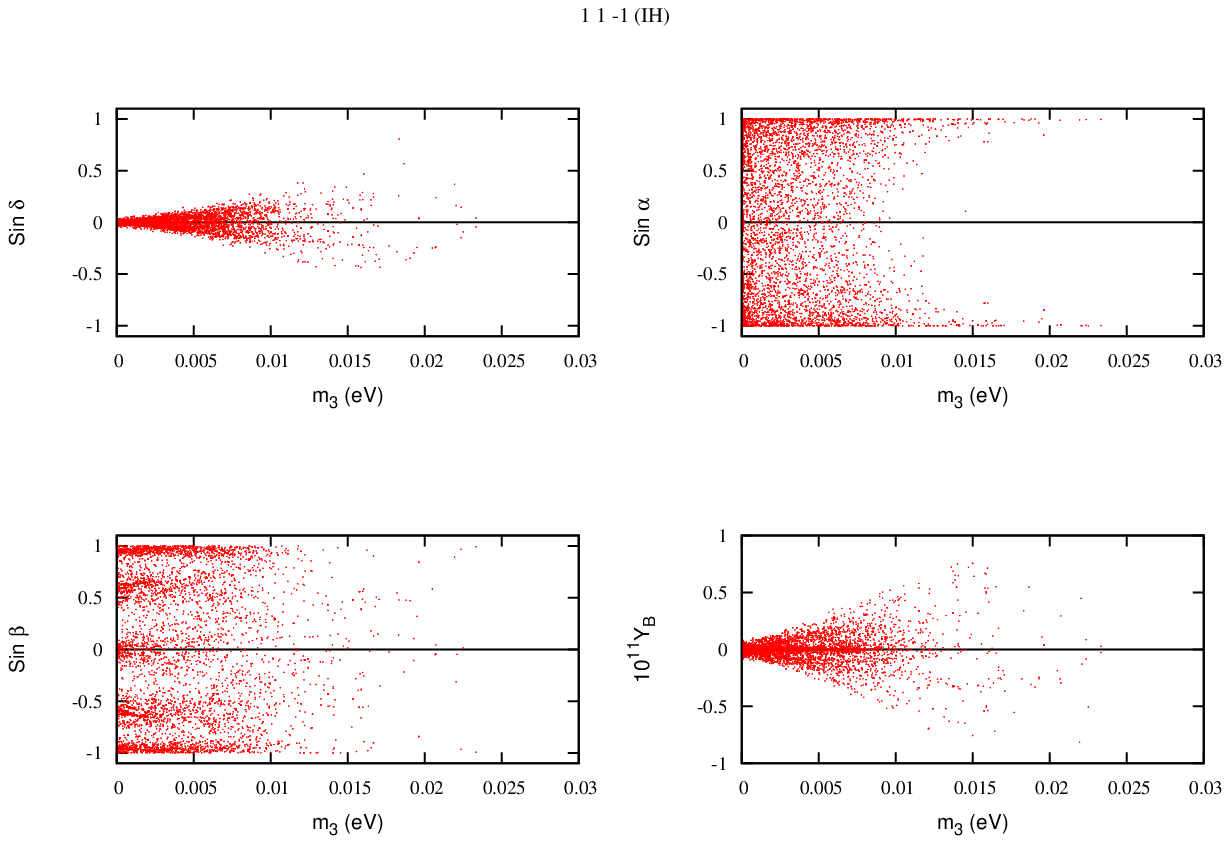} 
 \\
\end{array}$
 \caption{Inverted Hierarchy with triplet flavon vev alignment $(1,1,-1)$}
  \label{fig22}
\end{figure}
\begin{figure}[p]
\centering
$
\begin{array}{ccc}

\includegraphics[width=0.9\textwidth]{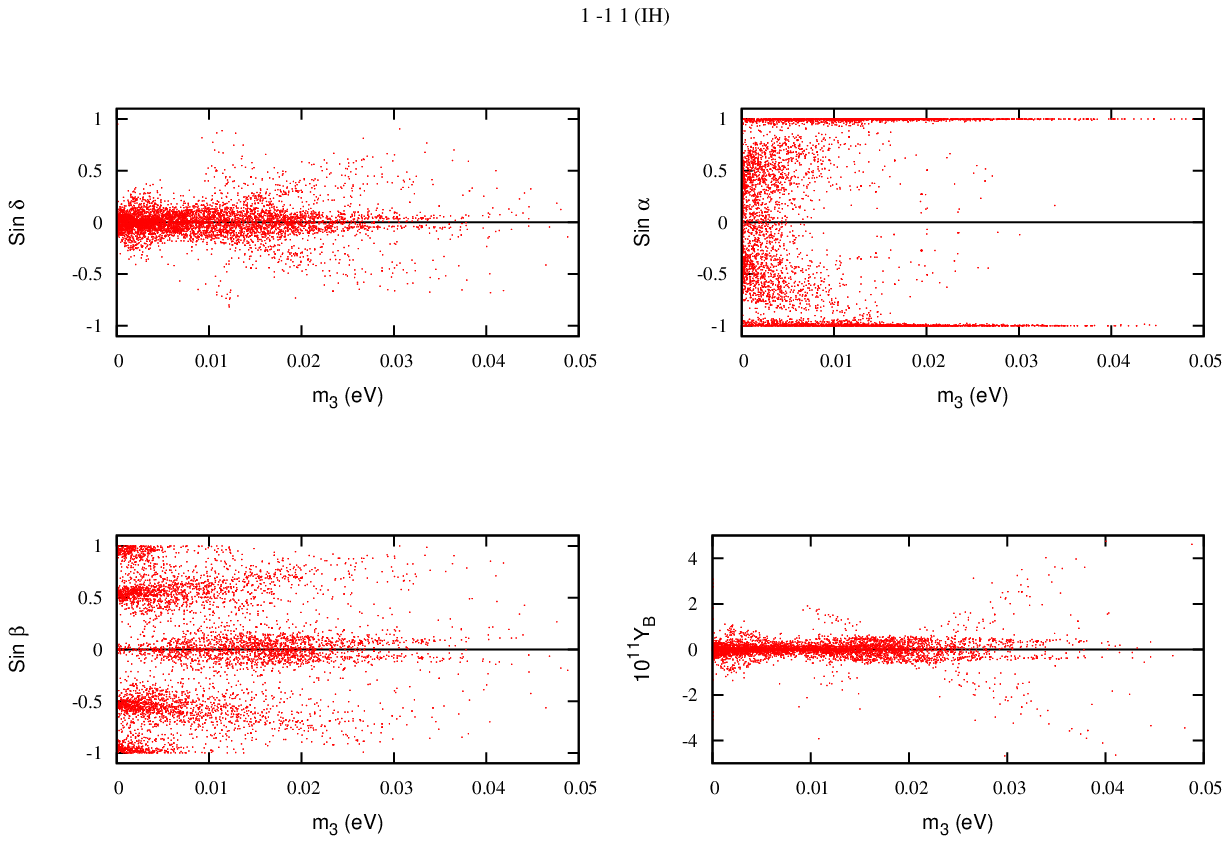} 
 \\
\end{array}$
 \caption{Inverted Hierarchy with triplet flavon vev alignment $(1,-1,1)$}
  \label{fig23}
\end{figure}
\begin{figure}[p]
\centering
$
\begin{array}{ccc}

\includegraphics[width=0.9\textwidth]{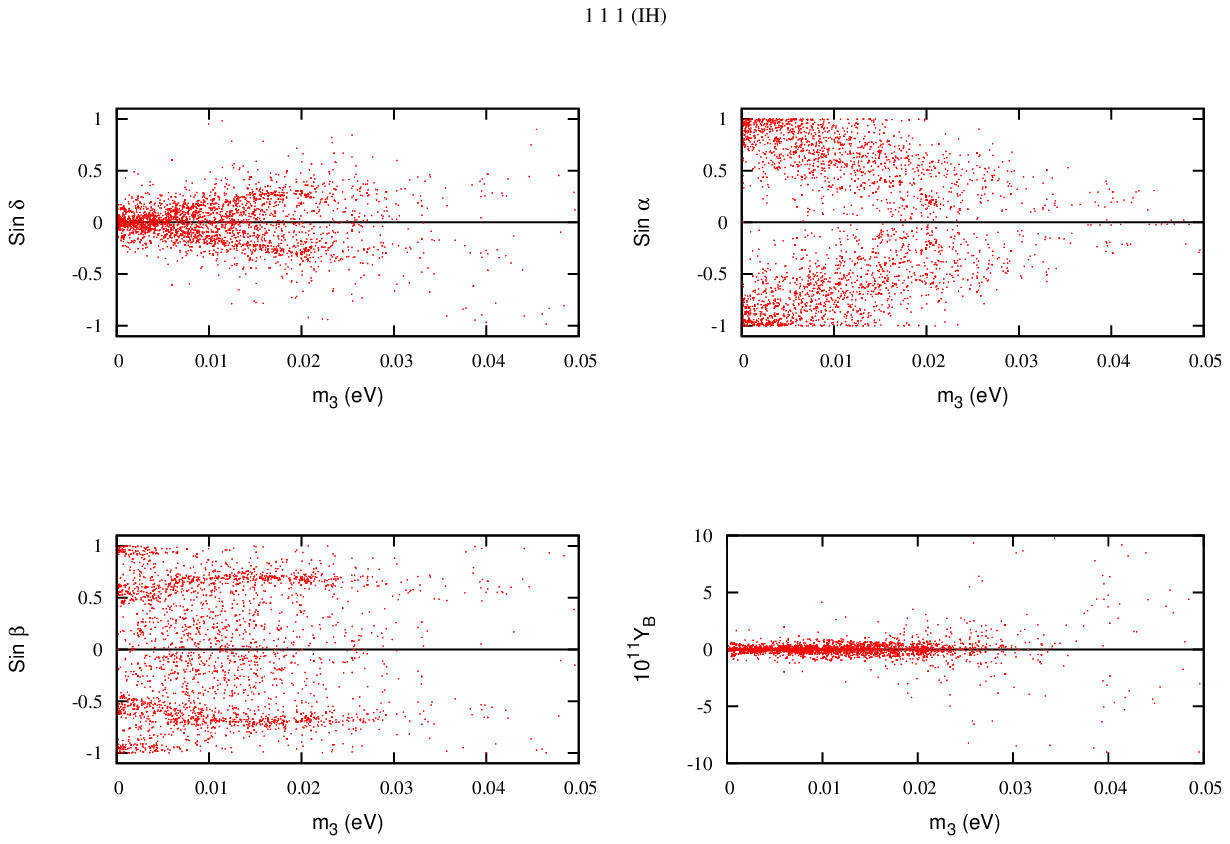} 
 \\
\end{array}$
 \caption{Inverted Hierarchy with triplet flavon vev alignment $(1,1,1)$}
  \label{fig24}
\end{figure}
\clearpage
\begin{figure}[p]
\centering
$
\begin{array}{ccc}

\includegraphics[width=0.9\textwidth]{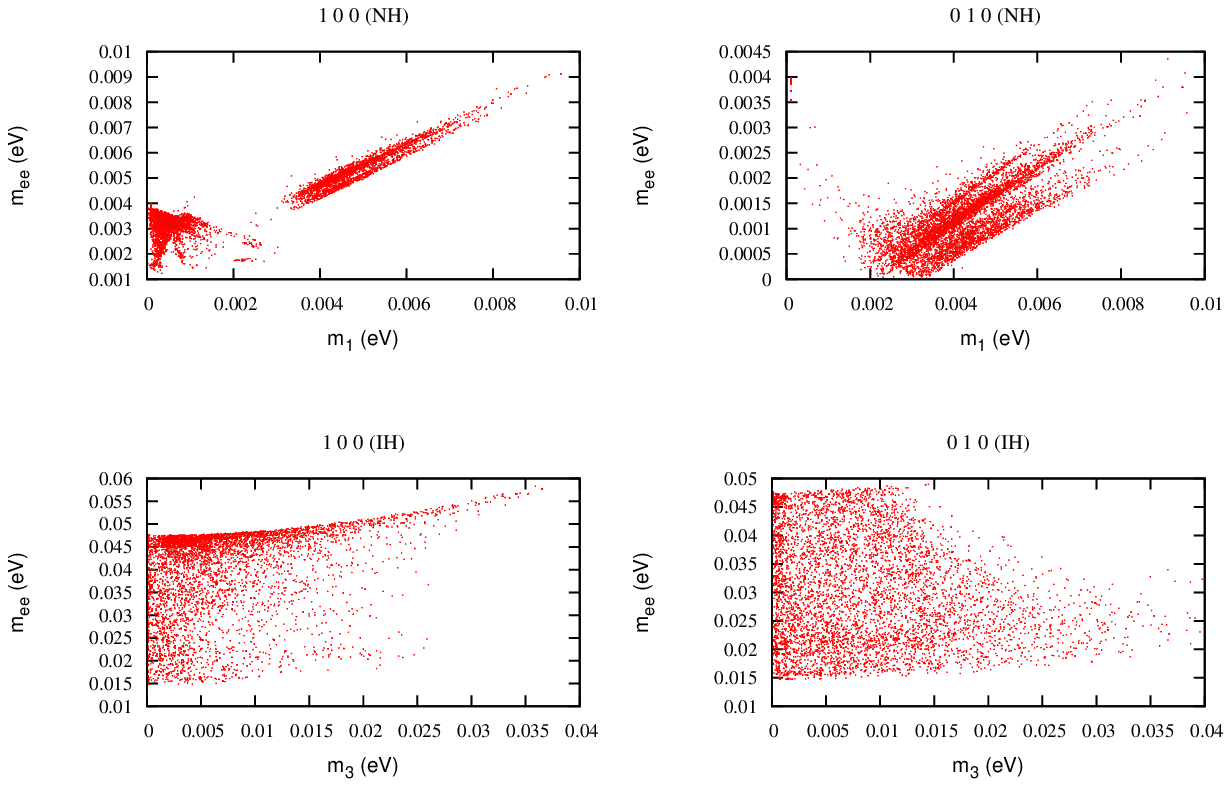} 
 \\
\end{array}$
 \caption{Effective neutrino mass for triplet flavon alignments $(1,0,0)$ and $(0,1,0)$}
  \label{fig35}
\end{figure}

\begin{figure}[p]
\centering
$
\begin{array}{ccc}

\includegraphics[width=0.9\textwidth]{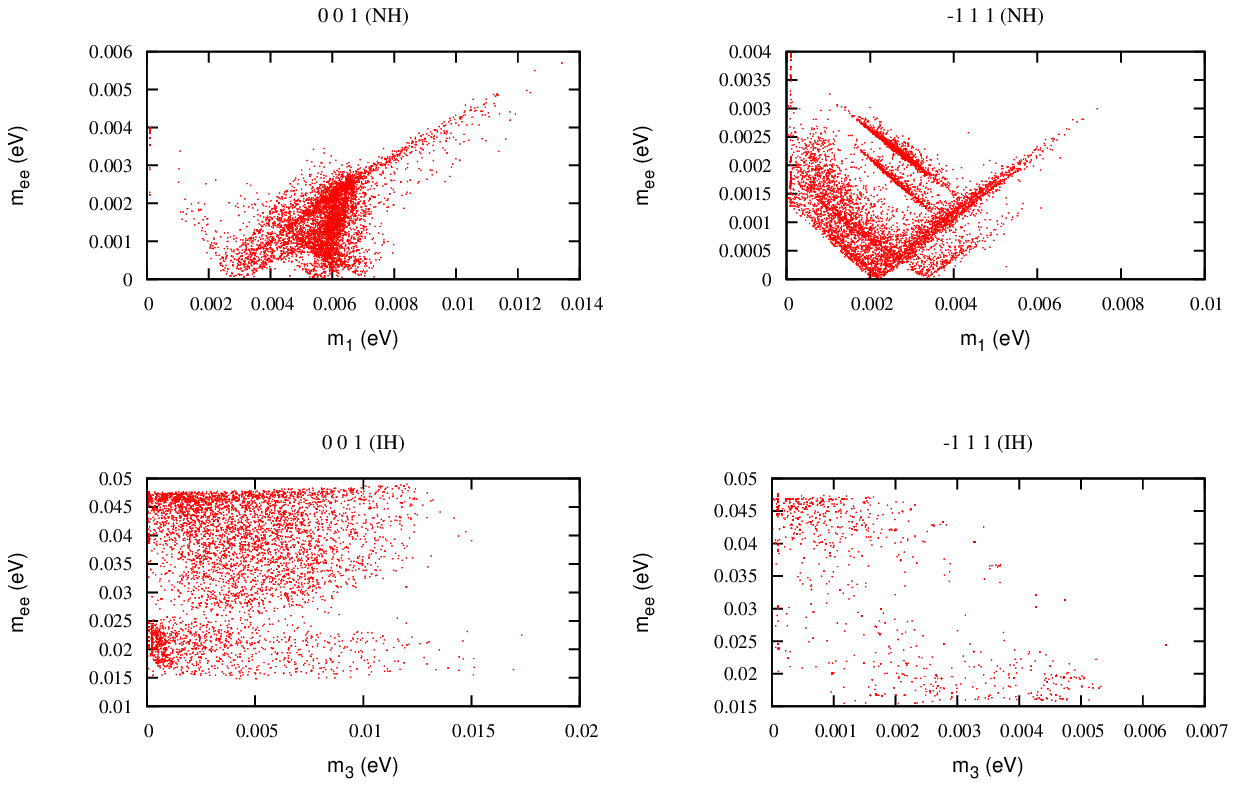} 
 \\
\end{array}$
 \caption{Effective neutrino mass for triplet flavon alignments $(0,0,1)$ and $(-1,1,1)$}
  \label{fig36}
\end{figure}
\begin{figure}[p]
\centering
$
\begin{array}{ccc}

\includegraphics[width=0.9\textwidth]{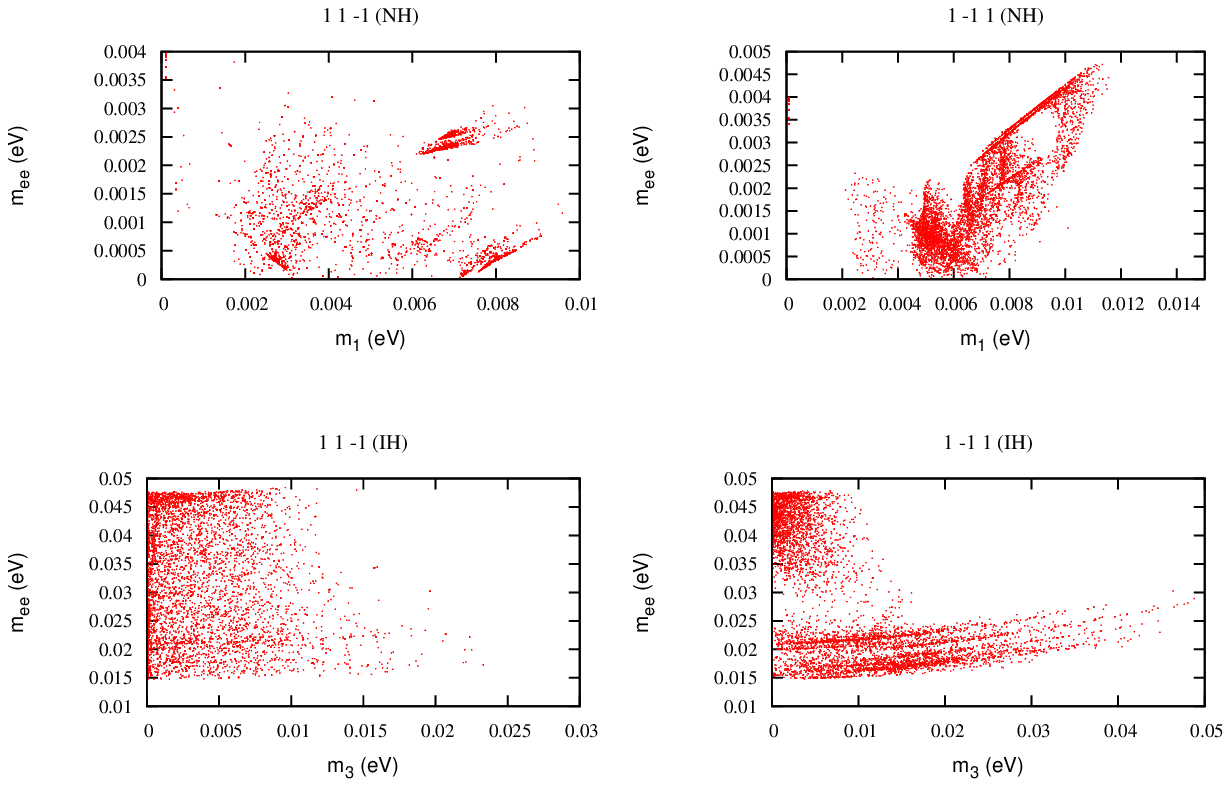} 
 \\
\end{array}$
 \caption{Effective neutrino mass for triplet flavon alignments $(1,1,-1)$ and $(1,-1,1)$}
  \label{fig37}
\end{figure}
\begin{figure}[p]
\centering
$
\begin{array}{ccc}

\includegraphics[width=0.9\textwidth]{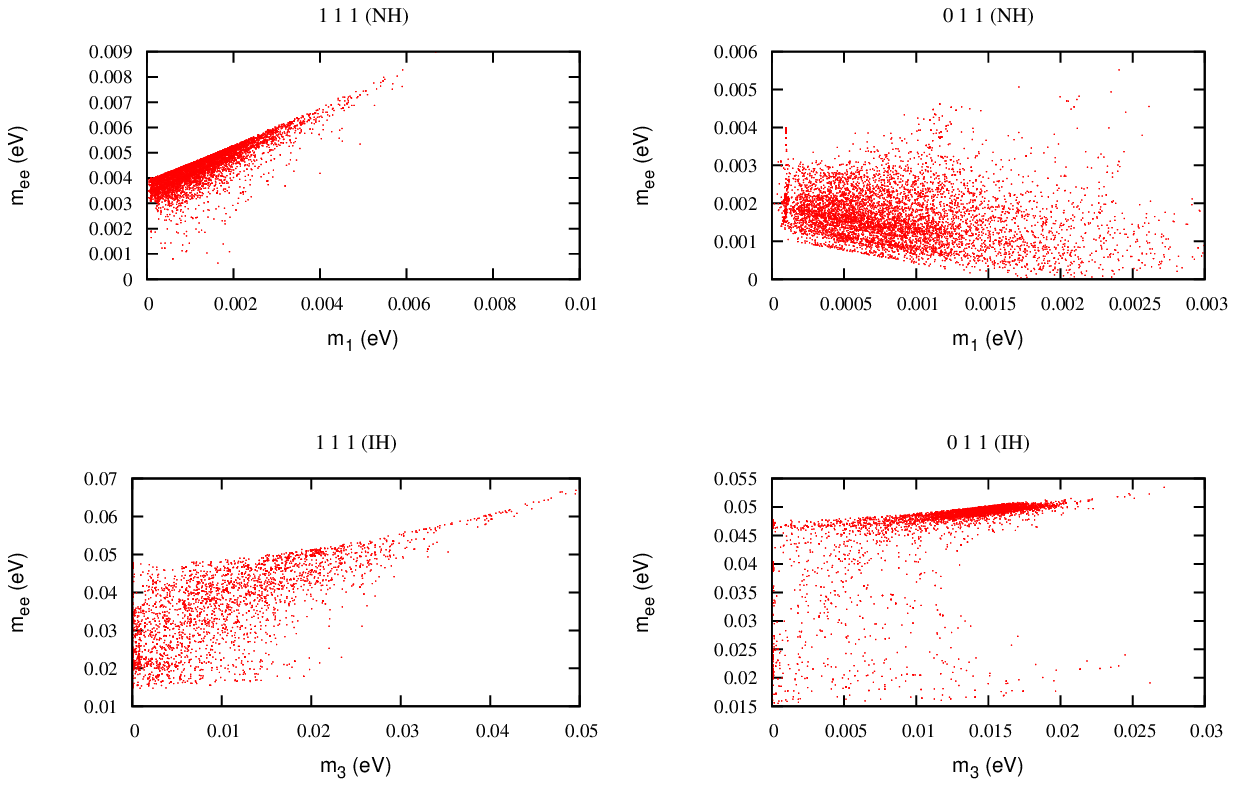} 
 \\
\end{array}$
 \caption{Effective neutrino mass for triplet flavon alignments $(1,1,1)$ and $(0,1,1)$}
  \label{fig38}
\end{figure}
\begin{figure}[p]
\centering
$
\begin{array}{ccc}

\includegraphics[width=0.9\textwidth]{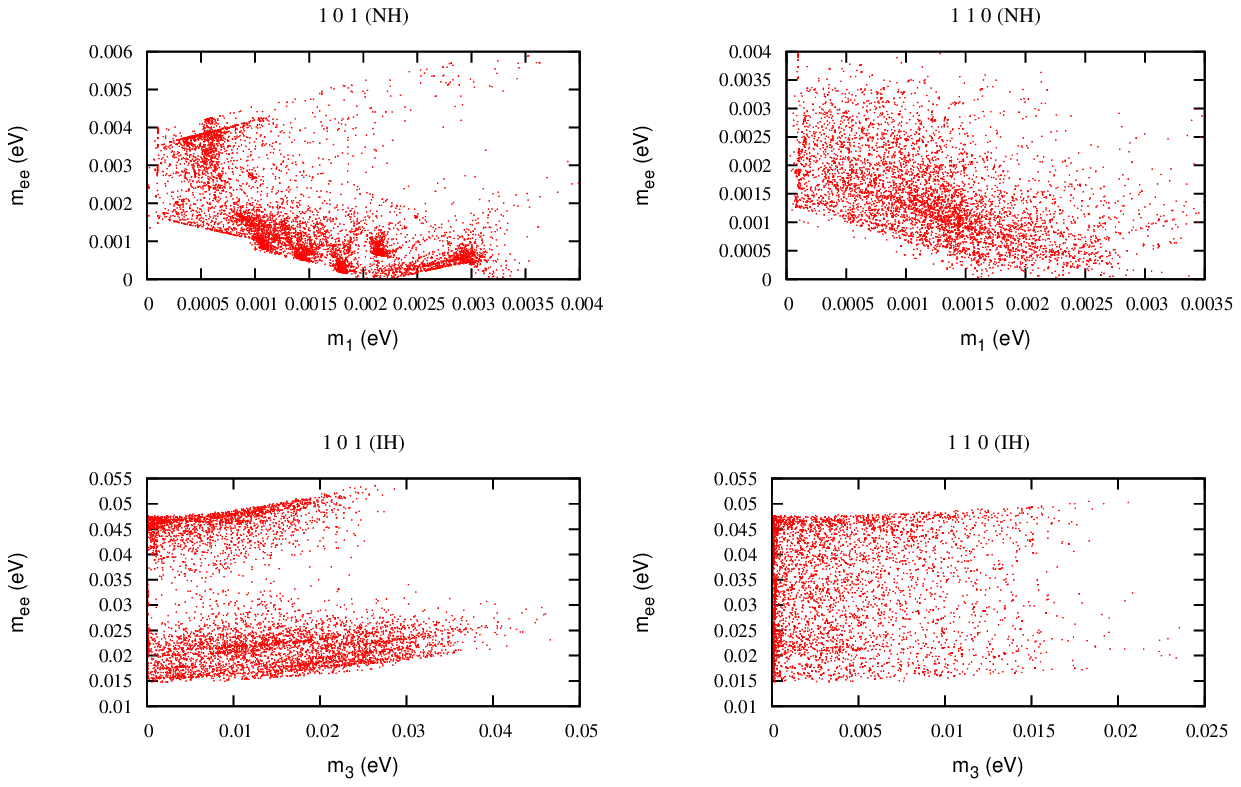} 
 \\
\end{array}$
 \caption{Effective neutrino mass for triplet flavon alignments $(1,0,1)$ and $(1,1,0)$}
  \label{fig39}
\end{figure}
\begin{figure}[p]
\centering
$
\begin{array}{ccc}

\includegraphics[width=0.9\textwidth]{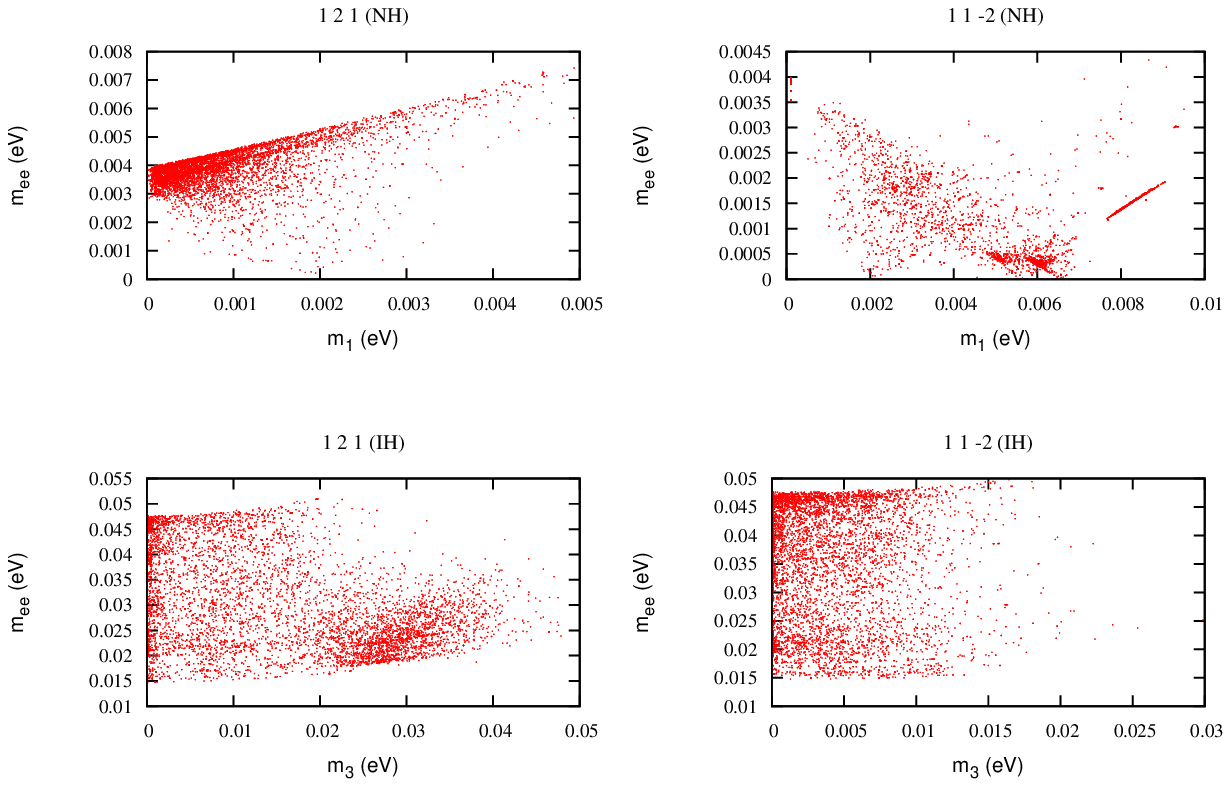} 
 \\
\end{array}$
 \caption{Effective neutrino mass for triplet flavon alignments $(1,2,1)$ and $(1,1,-2)$}
  \label{fig40}
\end{figure}
\begin{figure}[p]
\centering
$
\begin{array}{ccc}

\includegraphics[width=0.9\textwidth]{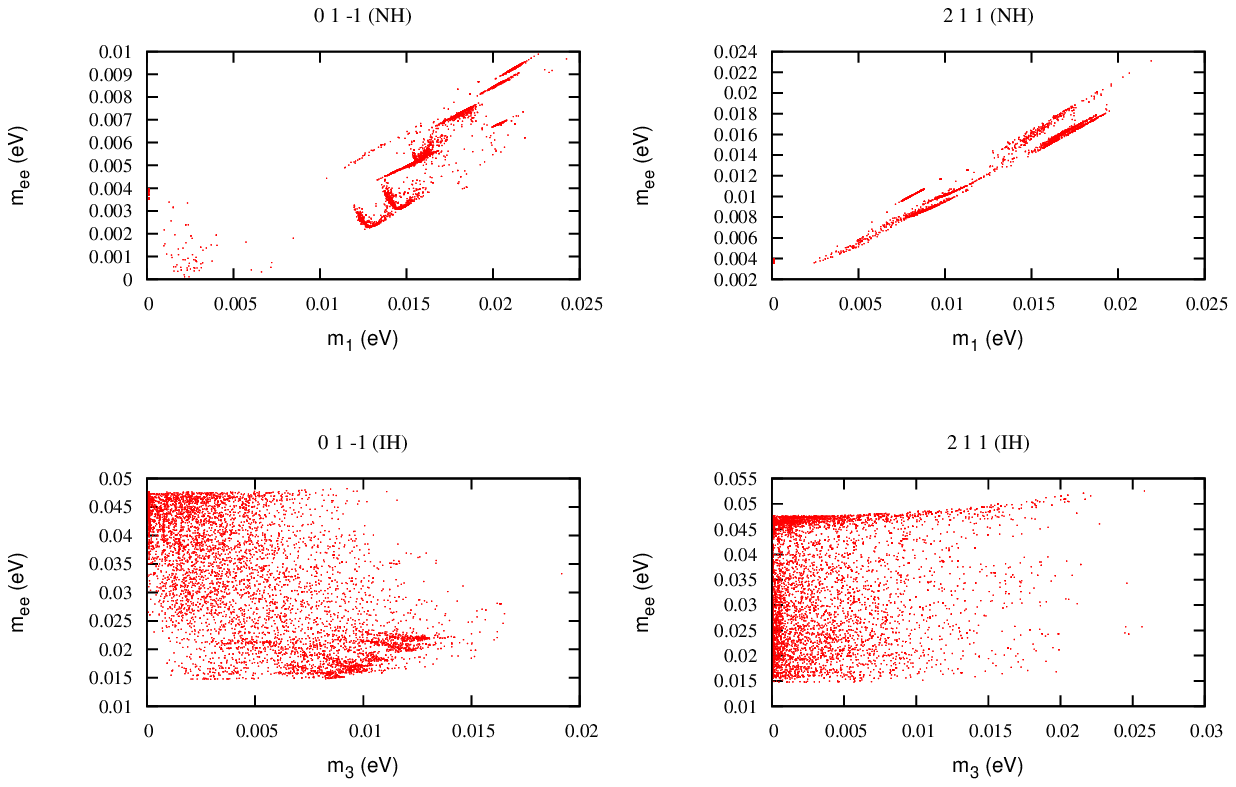} 
 \\
\end{array}$
 \caption{Effective neutrino mass for triplet flavon alignments $(0,1,-1)$ and $(2,1,1)$}
  \label{fig41}
\end{figure}
\begin{figure}[p]
\centering
$
\begin{array}{ccc}

\includegraphics[width=0.9\textwidth]{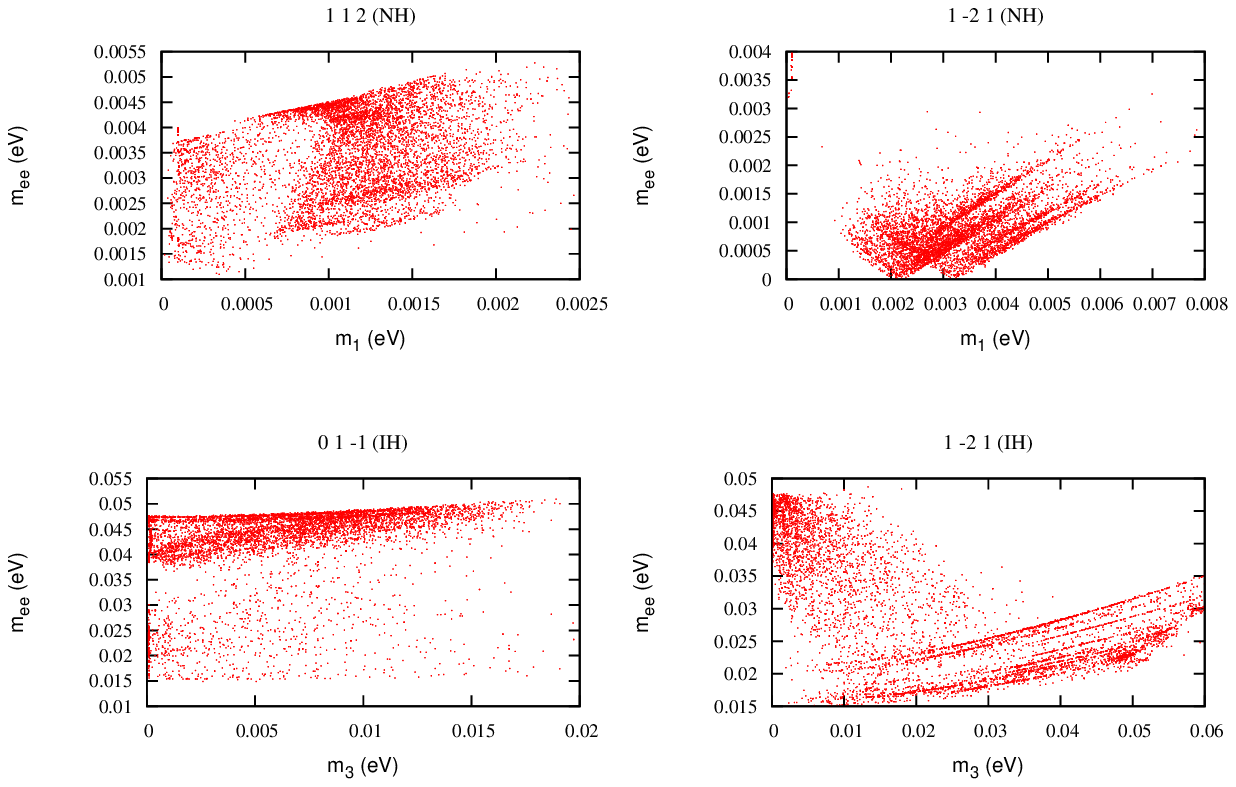} 
 \\
\end{array}$
 \caption{Effective neutrino mass for triplet flavon alignments $(1,1,2)$ and $(1,-2,1)$}
  \label{fig42}
\end{figure}
\begin{figure}[p]
\centering
$
\begin{array}{ccc}

\includegraphics[width=0.9\textwidth]{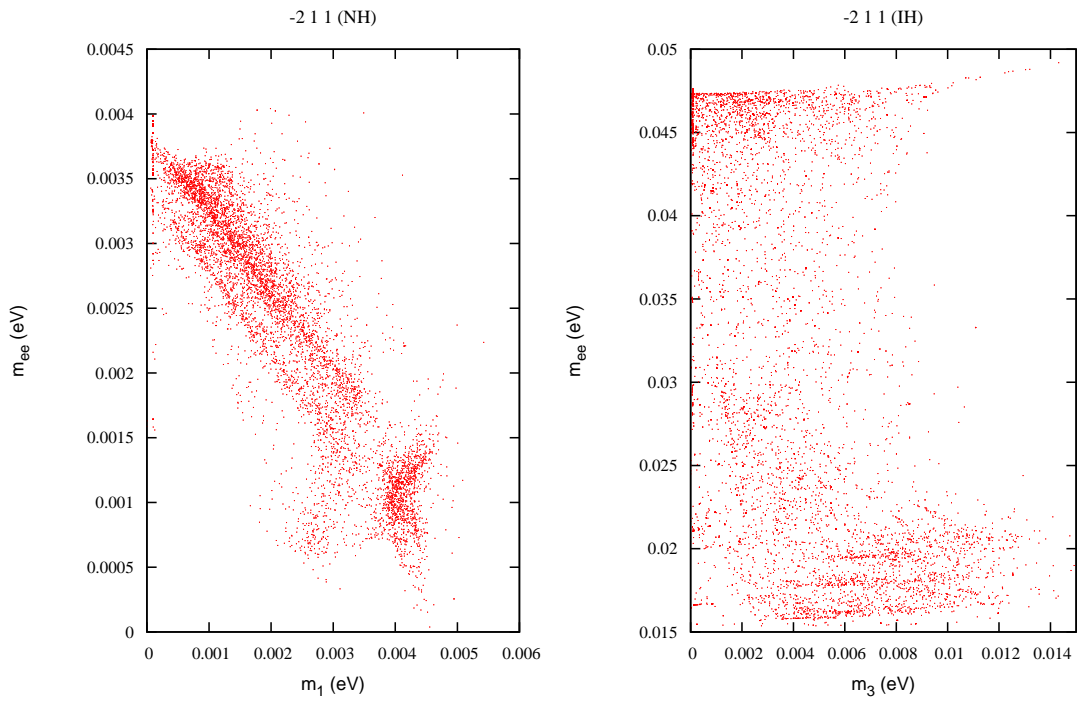} 
 \\
\end{array}$
 \caption{Effective neutrino mass for triplet flavon alignment $(-2,1,1)$}
  \label{fig43}
\end{figure}
\begin{table}[h]
\begin{tabular}{|l|l|l|l|l|l|l|l|}
\hline
\multicolumn{2}{|l|}{} & \multicolumn{2}{l|}{\bf{1 Flavor}} & \multicolumn{2}{l|}{\bf{2 Flavor}} & \multicolumn{2}{l|}{\bf{3 Flavor}} \\ \hline\hline
\multicolumn{2}{|l|}{$(\phi_a,\phi_b,\phi_c)$} &   NH       & IH          &   NH         &   IH        &     NH       &   IH        \\ \hline
\multicolumn{2}{|l|}{(1,0,0)} &      $\checkmark$     &  $\checkmark$           &    $\times$        &   $\times$         &       $\times$     &   $\times$         \\ 
\multicolumn{2}{|l|}{(0,1,0)} &        $\times$    &     $\checkmark$         &   $\times$         &    $\checkmark$        &    $\times$        &   $\times$         \\ 
\multicolumn{2}{|l|}{(0,0,1)} &     $\checkmark$       &    $\checkmark$         &  $\times$          &   $\times$         &       $\times$     &     $\times$       \\ \hline\hline
\multicolumn{2}{|l|}{(-1,1,1)} &   $\checkmark$         &     $\checkmark$         &   $\times$         & $\checkmark$           &    $\times$        &   $\times$         \\
\multicolumn{2}{|l|}{(1,1,-1)} &     $\checkmark$       &    $\checkmark$        &   $\times$         &  $\times$          &    $\times$        &   $\times$         \\ 
\multicolumn{2}{|l|}{(1,-1,1)} &    $\checkmark$        &     $\checkmark$         &   $\times$         &    $\times$        &   $\times$         &   $\times$         \\ 
\multicolumn{2}{|l|}{(1,1,1)} &    $\times$        &      $\checkmark$        &    $\times$        &    $\checkmark$       &    $\times$        &   $\times$         \\ \hline\hline
\multicolumn{2}{|l|}{(0,1,1)} &    $\times$        &      $\checkmark$        &    $\times$        &   $\checkmark$         &   $\times$         &  $\times$          \\ \hline
\multicolumn{2}{|l|}{(1,0,1)} &    $\checkmark$        &    $\checkmark$        &   $\times$         &   $\times$         &   $\times$         &   $\times$         \\ \hline
\multicolumn{2}{|l|}{(1,1,0)} &    $\times$        &     $\checkmark$        &  $\times$          &   $\times$         &   $\times$         &   $\times$         \\ \hline
\multicolumn{2}{|l|}{(0,1,-1)} &    $\times$        &     $\checkmark$        &  $\checkmark$          &  $\checkmark$         &   $\times$         &   $\times$         \\ \hline
\multicolumn{2}{|l|}{(2,1,1)} &    $\times$        &     $\checkmark$         &    $\checkmark$        &   $\times$         &    $\times$        &    $\times$        \\ \hline
\multicolumn{2}{|l|}{(1,1,2)} &    $\checkmark$        &     $\checkmark$         &   $\times$         &   $\times$         &    $\times$        &    $\times$        \\ \hline
\multicolumn{2}{|l|}{(1,2,1)} &   $\times$         &     $\checkmark$        &   $\times$         &  $\checkmark$         &    $\times$        &    $\times$        \\ \hline
\multicolumn{2}{|l|}{(1,-2,1)} &   $\times$         &    $\checkmark$         &   $\times$         &  $\checkmark$          &    $\times$        &   $\times$         \\ \hline
\multicolumn{2}{|l|}{(1,1,-2)} &    $\checkmark$        &    $\checkmark$         &   $\times$         &   $\times$         &    $\times$        &    $\times$        \\ \hline
\multicolumn{2}{|l|}{(-2,1,1)} &   $\checkmark$        &    $\checkmark$        &  $\times$          &    $\times$        &    $\times$        &    $\times$        \\ \hline
\end{tabular}
\caption{Summary of Baryon Asymmetry Results}
\label{tableYb1}
\end{table}
\begin{table}[h]
\begin{tabular}{|l|l|l|l|l|l|l|l|}
\hline
\multicolumn{2}{|l|}{} & \multicolumn{2}{l|}{\bf{1 Flavor}} & \multicolumn{2}{l|}{\bf{2 Flavor}}  \\ \hline\hline
\multicolumn{2}{|l|}{$(\phi_a,\phi_b,\phi_c)$} &   NH ($m_1 (\text{eV}) $)       & IH ($m_3 (\text{eV})$)         &   NH        ($m_1 (\text{eV})$) &   IH   ($m_3 (\text{eV})$)            \\ \hline
\multicolumn{2}{|l|}{(1,0,0)} &      $0.0059 $     &  $0.001-0.018$           &    $--$        &   $--$                \\ 
\multicolumn{2}{|l|}{(0,1,0)} &        $--$    &     $0.001-0.01$         &   $--$         &    $0.019,0.025$               \\ 
\multicolumn{2}{|l|}{(0,0,1)} &     $0.004,0.0053-0.0061$       &    $0.004-0.008$         &  $--$          &   $--$                 \\ \hline\hline
\multicolumn{2}{|l|}{(-1,1,1)} &   $0.0032,0.0032,0.0037$         &     $0.00165,0.0026$         &   $--$         & $0.0001$                  \\
\multicolumn{2}{|l|}{(1,1,-1)} &     $0.0025-0.0037,0.0054-0.0062$       &    $0.007,0.009$        &   $--$         &  $--$                 \\ 
\multicolumn{2}{|l|}{(1,-1,1)} &    $0.0057-0.0064$        &    $0.008-0.013$        &    $--$        &    $--$              \\ 
\multicolumn{2}{|l|}{(1,1,1)} &    $--$        &      $0.0013,0.005$       &   $--$        &    $0.039$               \\ \hline\hline
\multicolumn{2}{|l|}{(0,1,1)} &    $--$        &      $0.004,0.009,0.014$         &    $--$        &   $0.02$            \\ \hline
\multicolumn{2}{|l|}{(1,0,1)} &    $0.00025,0.00215$        &    $0.012-0.02 $        &   $--$         &   $--$                 \\ \hline
\multicolumn{2}{|l|}{(1,1,0)} &    $--$        &     $0.0072,0.009$        &  $--$          &   $--$                 \\ \hline
\multicolumn{2}{|l|}{(0,1,-1)} &    $--$        &     $0.003,0.012$        & $0.012-0.017$          &  $0.008-0.01$                \\ \hline
\multicolumn{2}{|l|}{(2,1,1)} &    $--$        &     $0.0051,0.0121$         &    $0.017,0.018$       &   $--$               \\ \hline
\multicolumn{2}{|l|}{(1,1,2)} &    $0.02 $        &     $0.002-0.009$         &   $--$         &   $--$               \\ \hline
\multicolumn{2}{|l|}{(1,2,1)} &   $--$         &     $0.007-0.013$        &   $--$         &   $0.025,0.045$              \\ \hline
\multicolumn{2}{|l|}{(1,-2,1)} &   $--$         &    $0.008,0.033$         &   $--$         &  $0.03-0.055$                \\ \hline
\multicolumn{2}{|l|}{(1,1,-2)} &    $0.032 $        &    $0.0028-0.0083$         &   $--$         &   $--$                \\ \hline
\multicolumn{2}{|l|}{(-2,1,1)} &    $0.003-0.006$        &    $0.0012-0.004$        &  $--$          &    $--$               \\ \hline
\end{tabular}
\caption{Values of lightest neutrino masses giving rise to correct baryon asymmetry }
\label{tableYb2}
\end{table}
\section{Results and Conclusion}
\label{sec:conclude}
We have studied a simple beyond standard model framework with type I seesaw mechanism of light neutrino masses where the symmetry of the standard model is enhanced by an additional $A_4$ flavor symmetry. The structures of leptonic mass matrices are dictated by the underlying $A_4$ flavor symmetry. We choose the $A_4$ flavon fields in such a way that the charged lepton mass matrix remains diagonal and the Dirac neutrino mass matrix takes the form shown in equation \eqref{mdirac}. The right handed neutrino mass matrix shown in equation \eqref{mright} is constructed by taking contribution from both singlet and triplet flavon fields under $A_4$. We also construct the right handed neutrino mass matrix by inverting the type I seesaw formula where the light neutrino mass matrix is evaluated in terms of the light neutrino parameters. Comparing these two mass matrices allow us to express the $A_4$ flavon vev's in terms of the light neutrino parameters. Out of the light neutrino parameters, we use the $3\sigma$ global fit values of two mass squared differences and three mixing angles, leaving four free parameters: the lightest neutrino mass, one Dirac CP phase and two Majorana CP phases. Instead of working in this four dimensional free parameter space, we evaluate them numerically by assuming the $A_4$ triplet flavon vev alignments to take some specific form which can preserve a $Z_2$ or $Z_3$ subgroup of $A_4$. We also consider alignments which break these subgroups for illustrative purposes. Evaluating these free parameters in the light neutrino sector also allows us to determine all the $A_4$ parameters numerically. We then use these parameters to calculate the baryon asymmetry through leptogenesis. The appropriate flavor regime of leptogenesis is decided by the strength of Dirac neutrino Yukawa coupling which is chosen by hand. For example, we show the parameter space of light neutrino parameters which satisfy a specific $Z_2$ or $Z_3$ preserving $A_4$ vacuum alignment in the figures starting from \ref{fig1} to \ref{fig24}. Each point shown in the first three panels of these figures correspond to those values of light neutrino free parameters $(m_{\text{lightest}}, \delta, \alpha, \beta)$ which give rise to one of the specific $A_4$ vev alignments listed in table \ref{vevA4}. The fourth panel in these figures show the baryon asymmetry in the two flavor regime for the region of parameter space shown in the first three panels. We also calculate the baryon asymmetry in one and three flavor regime, the summary of which can be found in table \ref{tableYb1}. It can be seen from table \ref{tableYb1} that none of the models give correct baryon asymmetry in the three flavor regime. This puts a constraint on the Dirac neutrino Yukawa coupling whose strength decides the flavor regime of leptogenesis in the models we are studying. Referring to the table \ref{lambdaN}, one can put a lower bound  $\lambda^2_N \approx 1.5 \times 10^{-5}$ below which none of the models discussed in this work give rise to successful thermal leptogenesis. In the two flavor regime where $\lambda^2_N \approx 10^{-4}-10^{-2}$, successful leptogenesis is possible with normal hierarchy only for the models with $A_4$ triplet vev alignment $(0,1,-1)$ and $(2,1,1)$. For inverted hierarchy, seven models corresponding to vev alignment $(0,1,0), (-1,1,1), (1,1,1), (0,1,1), (0,1,-1), (1,2,1), (1,-2,1)$ give rise to correct baryon asymmetry. In the one flavor regime, where  Dirac neutrino Yukawa coupling is of order unity, all the models with inverted hierarchy give successful leptogenesis whereas half of the models with normal hierarchy give rise to the desired lepton asymmetry, as shown in table \ref{tableYb1}.

We also list the values of lightest neutrino masses in table \ref{tableYb2} for which successful leptogenesis can be obtained for the models mentioned in table \ref{tableYb1}. It can be seen that the lightest neutrino mass can vary from $10^{-4}$ eV to $0.055$ eV depending on the model. The corresponding values of CP phases can be obtained from the figures showing their variation with lightest neutrino mass. Thus we can not only constrain the $A_4$ vacuum alignment from the requirement of producing correct baryon asymmetry through leptogenesis, but can also restrict the free parameters in the light neutrino sector to certain range of values. Even if the observed baryon asymmetry could have a different origin not related to the neutrino sector, these models can go through serious scrutiny at those experiments sensitive to CP violating phases and absolute neutrino mass scale. To have an independent experimental probe of the parameter space studied in this work, we also calculate the effective neutrino mass $m_{ee}$ to which neutrinoless double beta decay experiments are sensitive to. We plot the effective neutrino mass as a function of lightest neutrino mass for the parameter space allowed by $A_4$ vacuum alignment in the figures starting from \ref{fig35} to \ref{fig43}. One can see that phase II of GERDA experiments in future with sensitivity all the way down to 58-74 meV should be able to rule out some region of the parameter space discussed in this work. It is interesting to see from the figures \ref{fig1}-\ref{fig43} as well the tables \ref{tableYb1}-\ref{tableYb2} that within the same unbroken subgroup of $A_4$ namely, $Z_2$ or $Z_3$, simple permutations of $A_4$ triplet vev's can give rise to significant changes in the neutrino parameters $(m_{\text{lightest}}, \delta, \alpha, \beta)$ as well as other pbservables like baryon asymmetry and effective neutrino mass.

A possible extension of this work could be to show explicitly the realizations of $A_4$ vacuum alignment discussed here by minimizing the full scalar potential. Another important aspect not considered in the present work is to study the effect of renormalization group evolution (RGE) on the low energy parameters in the light neutrino sector. Here we have used the low energy best fit values of light neutrino parameters to solve the flavon equations. Since the flavons acquire vev at a high energy scale, one should consider the evolution of light neutrino parameters under RGE from low energy to the high energy scale. In the present work we have ignored these effects and have used the same low energy values of neutrino parameters to solve flavon equations as well as to calculate baryon asymmetry. We leave a detailed calculation incorporating the effects of RGE to future work.

\clearpage

\appendix
\section{$A_4$ product rules}
\label{appen1}
$A_4$ is a discrete non-abelian group of even permutations of four objects. It is also the symmetry group of a tetrahedron. This group has four irreducible representations: three one-dimensional and one three dimensional which are denoted by $\bf{1}, \bf{1'}, \bf{1''}$ and $\bf{3}$ respectively. Their product rules are given as
$$ \bf{1} \otimes \bf{1} = \bf{1}$$
$$ \bf{1'}\otimes \bf{1'} = \bf{1''}$$
$$ \bf{1'} \otimes \bf{1''} = \bf{1} $$
$$ \bf{1''} \otimes \bf{1''} = \bf{1'}$$
$$ \bf{3} \otimes \bf{3} = \bf{1} \otimes \bf{1'} \otimes \bf{1''} \otimes \bf{3}_a \otimes \bf{3}_s $$
where $a$ and $s$ in the subscript corresponds to anti-symmetric and symmetric parts respectively. Denoting two triplets as $(a_1, b_1, c_1)$ and $(a_2, b_2, c_2)$ respectively, their direct product can be decomposed into the direct sum mentioned above as
$$ \bf{1} \backsim a_1a_2+b_1c_2+c_1b_2$$
$$ \bf{1'} \backsim c_1c_2+a_1b_2+b_1a_2$$
$$ \bf{1''} \backsim b_1b_2+c_1a_2+a_1c_2$$
$$\bf{3}_s \backsim (2a_1a_2-b_1c_2-c_1b_2, 2c_1c_2-a_1b_2-b_1a_2, 2b_1b_2-a_1c_2-c_1a_2)$$
$$ \bf{3}_a \backsim (b_1c_2-c_1b_2, a_1b_2-b_1a_2, c_1a_2-a_1c_2)$$

\section{$A_4$ Flavon vev's}
\label{appen2}
\begin{equation}
\begin{split}
\phi_{a}=\frac{F}{m_1}\left(c^2_{12}c^2_{13}-(e^{i\delta}c_{12}c_{23}s_{13}-s_{12}s_{23})(c_{23}s_{12}+e^{i\delta}c_{12}s_{13}s_{23})\right)\\+\frac{F}{m_2}e^{2i\alpha}\left(c^2_{13}s^2_{12}+(e^{i\delta}c_{23}s_{12}s_{13}+c_{12}s_{23})(c_{12}c_{23}-e^{i\delta}s_{12}s_{13}s_{23})\right)\\
+\frac{F}{m_3}e^{2i\beta}\left(s^2_{13}-e^{2i\delta}c^2_{13}c_{23}s_{23}\right) 
\end{split}
\label{eq1appen2}
\end{equation}
\begin{equation}
\begin{split}
\phi_b=\frac{F}{m_1}\left((e^{i\delta}c_{12}c_{23}s_{13}-s_{12}s_{23})^2+c_{12}c_{13}(c_{23}s_{12}+e^{i\delta}c_{12}s_{13}s_{23})\right)\\\frac{F}{m_2}e^{2i\alpha}\left((e^{i\delta}c_{23}s_{12}s_{13}+c_{12}s_{23})^2+(c_{13}s_{12}(-c_{12}c_{23}+e^{i\delta}s_{12}s_{13}s_{23}))\right)\\+ \frac{F}{m_3}\left(e^{2i(\beta+\delta)}c^2_{13}c^2_{23}-e^{i(2\beta+\delta)}c_{13}s_{13}s_{23}\right)
\end{split}
\label{eq2appen2}
\end{equation}

\begin{equation}
\begin{split}
\phi_c=\frac{F}{m_1}\left((e^{i\delta}c_{12}s_{13}s_{23}+c_{23}s_{12})^2+c_{12}c_{13}(e^{i\delta}c_{12}c_{23}s_{13}-s_{12}s_{23})\right)\\\frac{F}{m_2}e^{2i\alpha}\left((-e^{i\delta}s_{12}s_{13}s_{23}+c_{12}c_{23})^2+(c_{13}s_{12}(c_{12}s_{23}+e^{i\delta}c_{23}s_{12}s_{13}))\right)\\
\frac{F}{m_3}\left(e^{2i(\beta+\delta)}c^2_{13}s^2_{23}-e^{i(2\beta+\delta)}c_{13}c_{23}s_{13}\right)
\end{split}
\label{eq3appen2}
\end{equation}

\begin{equation}
\begin{split}
\eta=\frac{F}{3m_1m_2m_3}\left(e^{2i\beta}m_1m_2s^2_{13}-2m_2m_3c_{23}s^2_{12}s_{23}+2e^{2i(\alpha+\delta)}m_1m_3c_{23}s^2_{12}s^2_{13}s_{23}\right)\\+\frac{F}{3m_1m_2m_3}\left(m_1c^2_{13}(e^{2i\alpha}m_3s^2_{12}+2e^{2i(\beta+\delta)}m_2c_{23}s_{23})\right)+\\\frac{F}{3m_1m_2m_3}\left(m_3c^2_{12}\left(c^2_{13}m_2-2c_{23}s_{23}(e^{2i\alpha}m_1-e^{2i\delta}m_2s^2_{13})\right)\right)\\-\frac{F}{3m_1m_2m_3}\left(2e^{i\delta}m_3c_{12}s_{12}s_{13}(e^{2i\alpha}m_1-m_2)(c^2_{23}-s^2_{23})\right)
\end{split}
\label{eq4appen2}
\end{equation}

\begin{equation}
\begin{split}
\psi=\frac{F}{3m_1m_2m_3}\left(m_2m_3c^2_{23}s^2_{12}+2e^{i\delta}m_1c_{13}c_{23}s_{13}(e^{2i\beta}m_2-e^{2i\alpha}m_3s^2_{12})\right)-\\\frac{F}{3m_1m_2m_3}\left(2m_3c_{12}s_{12}s_{23}(e^{2i\alpha}m_1-m_2)(c_{13}+e^{i\delta}c_{23}s_{13})\right)\\+\frac{F}{3m_1m_2m_3}\left(e^{2i\delta}m_1s^2_{23}(e^{2i\beta}m_2c^2_{13}+e^{2i\alpha}m_3s^2_{12}s^2_{13})\right)\\+\frac{F}{3m_1m_2m_3}\left(m_3c^2_{12}(e^{2i\alpha}m_1c^2_{23}-2e^{2i\delta}m_{2}c_{13}c_{23}s_{13}+e^{2i\delta}m_2s^2_{13}s^2_{23})\right)
\end{split}
\label{eq5appen2}
\end{equation}

\begin{equation}
\begin{split}
\chi=\frac{F}{3m_1m_2m_3}\left(m_1m_2c^2_{13}c^2_{23}e^{2i(\beta+\delta)}+2m_3c_{12}c_{13}c_{23}s_{12}(e^{2i\alpha}m_1-m_2)\right)\\+\frac{F}{3m_1m_2m_3}\left(2c_{13}(-m_2m_3c^2_{12}s_{13}s_{23}e^{i\delta}+e^{i\delta}m_1s_{13}s_{23}(e^{2i\beta}m_2-e^{2i\alpha}m_3s^2_{12}))\right)\\+\frac{F}{3m_1m_2m_3}\left(2m_3c_{12}c_{23}s_{12}s_{13}s_{23}e^{i\delta}(e^{2i\alpha}m_1-m_2)+(m_3c^2_{12}e^{2i\delta}m_2c^2_{13}s^2_{13})\right)\\+\frac{F}{3m_1m_2m_3}\left(m_1m_3c^2_{12}s^2_{23}e^{2i\alpha}+m_3s^2_{12}(e^{2i(\alpha+\delta)}m_1c^2_{23}s^2_{13}+m_2s^2_{23})\right)
\end{split}
\label{eq6appen2}
\end{equation}
 where $F=\frac{v^2\lambda^2_{N}}{\Lambda_{RR}}$

\end{document}